\documentclass[sigconf]{acmart}

\usepackage{booktabs} 

\usepackage{stfloats}
\usepackage{graphicx}
\usepackage{amssymb}
\usepackage{subfigure}
\usepackage{indentfirst}
\usepackage{multirow}
\usepackage{algpseudocode}
\usepackage{algorithmicx}
\usepackage{algorithm}
\usepackage{amsfonts}
\usepackage{mathrsfs}
\usepackage{amsmath}
\usepackage{balance}
\usepackage{breakurl}
\newcommand{\stitle}[1]{\vspace{1ex}{\bf #1}}
\newcommand{\eat}[1]{}

\copyrightyear{2019}
\acmYear{2019}
\setcopyright{iw3c2w3}
\acmConference[WWW '19]{Proceedings of the 2019 World Wide Web Conference}{May 13--17, 2019}{San Francisco, CA, USA}
\acmBooktitle{Proceedings of the 2019 World Wide Web Conference (WWW '19), May 13--17, 2019, San Francisco, CA, USA}
\acmPrice{}
\acmDOI{10.1145/3308558.3313568}
\acmISBN{978-1-4503-6674-8/19/05} 

\begin{document}
\title{Personalized Bundle List Recommendation}

\author{
	Jinze Bai\textsuperscript{1},
	Chang Zhou\textsuperscript{2*},
	Junshuai Song\textsuperscript{1},
	Xiaoru Qu\textsuperscript{1},
	Weiting An\textsuperscript{2},
	Zhao Li\textsuperscript{2},
	Jun Gao\textsuperscript{1}\\
	\small
	\textsuperscript{1}Key Laboratory of High Confidence Software Technologies, EECS, Peking University \,
	\textsuperscript{2}Alibaba Group \,
	\textsuperscript{*}Corresponding Author\\
	\{baijinze, songjs, quxiaoru, gaojun\}@pku.edu.cn,
	\{ericzhou.zc, weiting.awt, lizhao.lz\}@alibaba-inc.com
}

\renewcommand{\shortauthors}{J. Bai et al.}

\begin{abstract}
Product bundling, offering a combination of items to customers, is one of the marketing strategies commonly used in online e-commerce and offline retailers. A high-quality bundle generalizes frequent items of interest, and diversity across bundles boosts the user-experience and eventually increases transaction volume. In this paper, we formalize the personalized bundle list recommendation as a structured prediction problem and propose a bundle generation network (BGN), which decomposes the problem into quality/diversity parts by the determinantal point processes (DPPs). BGN uses a typical encoder-decoder framework with a proposed feature-aware softmax to alleviate the inadequate representation of traditional softmax, and integrates the masked beam search and DPP selection to produce high-quality and diversified bundle list with an appropriate bundle size. We conduct extensive experiments on three public datasets and one industrial dataset, including two generated from co-purchase records and the other two extracted from real-world online bundle services. BGN significantly outperforms the state-of-the-art methods in terms of quality, diversity and response time over all datasets. In particular, BGN improves the precision of the best competitors by 16\% on average while maintaining the highest diversity on four datasets, and yields a 3.85x improvement of response time over the best competitors in the bundle list recommendation problem.
\end{abstract}

%
%
\eat{

\ccsdesc[500]{Web Mining and Content Analysis~Recommendation systems}
\ccsdesc[300]{Web Mining and Content Analysis~Web Activities and Dynamics}
\ccsdesc{Computing methodologies~Machine learning}
}

\keywords{Bundle Recommendation; Bundle Generation; Diversity}
\fancyhead{}

\settopmatter{printacmref=false, printfolios=false}

\maketitle

{\fontsize{8pt}{8pt} \selectfont \noindent
\textbf{ACM Reference Format:}\\
Jinze Bai, Chang Zhou, Junshuai Song, Xiaoru Qu, Weiting An, Zhao Li, Jun Gao. 2019. Personalized Bundle List Recommendation. In \textit{Proceedings of the 2019 World Wide Web Conference (WWW’19), May 13--17, 2019, San Francisco, CA, USA.} ACM, New York, NY, USA, 11 pages. \url{https://doi.org/10.1145/3308558.3313568}}

\section{Introduction}
A bundle is a collection of products or services for sale as a whole, and the bundle list recommendation is to recommend a personalized list of bundles. Traditional business domains, e.g., communication services, offline retailers and supermarkets, take bundling as a critical marketing strategy, and they usually make bundles by human insights or non-personalized mining methods \cite{apriori}. Modern e-commerce websites and online service business, e.g., Amazon, Taobao, Steam and Netflix, also deploy new applications \cite{Pathak2017Generating,xie2014generating}, which recommend and sale a list of bundles rather than a list of items.

Customers, sellers, as well as the recommendation platform, could benefit from the personalized bundle recommendation service.
For customers, high-quality bundles broaden user interests and indicate the complementary products directly in the shortest path after purchase. It is well known in classic marketing research that carefully-designed bundle brings mutual promotions.

For sellers, bundling increases per customer transaction. The average order size for walmart.com is 2.3 \cite{zhu2014bundle}. By attaching some highly correlated products, the seller can launch more products and increase gross merchandise volume (GMV). Besides, buying a bundle of products may cost less than buying each product separately for both customers and sellers. For example, it is a common practice to waive shipping fee if the items in one order exceed a certain amount in e-commerce websites.

For the recommendation system, bundling is a more efficient way to organize and present products. The recommendation list is structured by bundles where the products are highly correlated in contrast to the list of items, and multiple diversified bundles avoid monotonous choices for users. The recommendation system shows bundle as a whole and thus presents more products per unit area. It is crucial to reduce swipe-up operations for improving user experiences for the mobile application.

\begin{figure*}[htbp]
	\centering
	\includegraphics[scale=0.74]{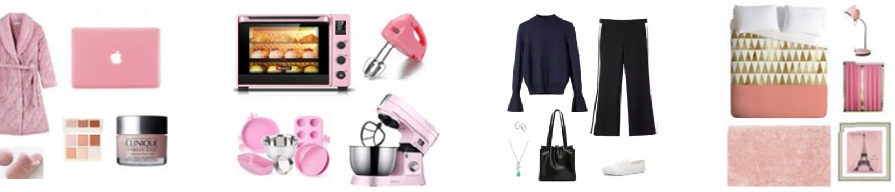}
	\caption{Bundle List Recommendation Example}
	\label{fig:bundle_example}
\end{figure*}

In Figure \ref{fig:bundle_example}, we illustrate a \eat{real-world} bundle list example. The items in the bundle should be highly correlated. Besides, diversity across bundles becomes a critical metric as an interpretable, business-oriented goal, which is independent of click-through rate (CTR) and precision, trying to avoid tedious choices and quantify the user-experience in bundle recommendation system. The identical high-quality item tends to be generated in different bundles \cite{Pathak2017Generating}, which compose the confusing and monotonous bundle list, so there is a trade-off between precision/likelihood and diversity. Other meaningful problems, e.g., how to set the price discount for maximizing the profit, how to compose the picture of the bundle, are beyond the scope of this paper and can be the future work. Our paper focuses on how to generate high-quality and diversified bundles.

There are several challenges for the personalized bundle recommendation problem. First, the recommendation system should generate high-quality bundles automatically and flexibly without manual involvement. Frequent itemsets mining algorithms provide high-quality bundling but are not personalized. For modeling the distribution of bundle, we use the co-purchase data or pre-defined bundles to supervise the quality of bundles with the user context. However, the sequence generation methods suffer from inadequate representation for rich features in bundle recommendation scenario, because the weight matrix of traditional softmax only considers the item\_id. Besides, in practice, we may need the larger size of bundles for the purpose of sellers' promotion, but seq2seq usually suffers from generating short sequences \cite{koehn2017six}.

\eat{However, the sequence generation methods usually suffer from generating short sequences \cite{koehn2017six}, and in practice, we may need larger size of bundles for the purpose of sellers' promotion. The distribution of the co-purchase data may not be consistent with the expected distribution of the bundle size for the target application. We use the proposed masked beam search to adjust the size of the generated bundles.}

Second, diversity across bundles emerges as a critical metric for bundle recommendation. Different bundles with the same high-quality item may reach high ranking scores at the same time, so duplicated items tend to be seen in these generated bundle list \cite{Pathak2017Generating}. Besides duplication, similar items due to the similarity of features may produce confusing and tedious bundle list and decrease user interest. A bundle recommendation system should consider the cross-bundle diversity considering both duplication and similarity of items. Here, the diversity is a explicit metric instead of some hidden relationships of similarity learned from data. The explicit similarity pre-defined by side information and prior knowledge provides the exact signal for the purpose of measurement and adjustment. We use the determinantal point processes (DPPs) as the objective function which considers the global negative correlations during inference to decompose the quality and diversity.

Last, compared to the traditional item list recommendation, the search space is the doubly exponential number of items for the bundle recommendation, since a bundle can contain arbitrary-length combination theoretically and a list contains multiple bundles \cite{deng2013complexity}. Bundle sparsity should be carefully considered because the training data in the form of co-purchase contains few bundle ground truths w.r.t. the whole search space. Most of the bundles in search space are low-quality, so we need a method to generate high-quality candidate bundles efficiently for each user, instead of the ranking method applied in all the possible bundles.

In this paper, we aim to the problem of the personalized bundle list recommendation and propose a \textbf{Bundle Generation Network (BGN)}.
We summarize the contributions of this paper as follows:

\begin{itemize}
	
	\item It is the first attempt to address the bundle list recommendation problem by the sequence generation approach considering both the quality and diversity. We decompose the generation model based on determinantal point processes, and integrate the neural network with DPP selection to produce the high-quality and diversified bundle list.
	
	\item We propose a feature-aware softmax in bundle recommendation scenario, for making up the inadequate representation of traditional seq2seq for rich features, which improves the modeling of quality significantly. The loss of feature-aware softmax is a generalization of optimizing BPR loss at each step.
	
	\item We conduct extensive experiments on both the real-world dataset and public datasets. Our model improves the precision of the best baseline by 16\% on average while maintaining the highest diversity on four datasets, and shows the fastest response time in the bundle list recommendation problem.  Besides, BGN can control the bundle size with a tolerable reduction in precision.
	
\end{itemize}

\section{Related Works}
We present the existing bundle recommendation approaches and introduce the sequence generation model (Seq2seq) and Determinantal Point Processes (DPPs) as related works.

\subsection{Bundle Recommendation}
Product bundling is one of the marking strategies commonly used in both offline retailers and online e-Commerce websites. Frequent itemsets mining algorithms like Apriori \cite{apriori} provide high-quality bundling but are not personalized. Parameswaran  et al. \cite{Parameswaran2011Recommendation} study some complex constraints like cross-item dependencies in personalized curriculum bundling. The complexity of top-K bundle recommendation is discussed comprehensively in \cite{deng2013complexity}. All above only use frequency of items, which cannot utilize rich features and fail to generalize on similar items and users.

For the single bundle recommendation, Zhu et al. \cite{zhu2014bundle} propose a personalized method based on integer quadratic programming. This method considers the first and the second terms of cross-item dependencies. However, the estimation of the second term is heuristic, and this method ignores the higher order terms of cross-item dependencies, which gives a sub-optimal result. The bundle recommendation problem to groups rather than users is investigated by \cite{qi2016recommending}. For the bundle list recommendation, Xie et al. \cite{xie2014generating} generate top-k bundles via sampling and preference elicitation. This method uses an additive utility function with a linear combination of the corresponding (aggregate) feature values, which fails to capture the complicated dependencies. In \cite{Pathak2017Generating}, Pathak and Gupta build an optimization criterion upon the latent factor model, Bayesian Personalized Ranking (BPR) \cite{bpr}, with the mean pair-wise Pearson correlation of the items. This method generates personalized bundles by a two-stage approach, combining the corresponding preference scores with a greedy annealing schedule, which might be caught in the local optimum due to randomness and instability of annealing.

\subsection{Sequence Generation Approaches}
In recent years, sequence generation methods achieve great successes in many tasks like Machine Translation \cite{seq2seq} and Dialog Generation \cite{dialog_gen}. They adopt an encoder-decoder architecture to generate sequences, and use the beam search strategy \cite{wiseman2016sequence} to produce the maximum possible sequences approximately.

In recommendation systems, some works based on RNN, CNN, attention extract the user embedding from the behaviors for each user \cite{wu2016personal,cnn_wsdm,atrank}. The recurrent model and the memory network are also adopted for sequential recommendation \cite{liu2016context,chen2018sequential}. The difference between the sequential recommendation and the bundle recommendation is that, the former uses the encoder only, while the latter uses the encoder to model the user context and uses the decoder to model the bundle distribution. The decoder usually has separate trainable parameters without sharing with the encoder. Besides, The bundle recommendation produces multiple bundles, so we need to know when to end the generation and control diversity of bundles.

\subsection{Determinantal Point Processes}
Determinantal Point Processes (DPPs), first introduced to model fermion behavior,  have gained popularity due to their elegant balancing of quality and diversity. DPPs have been studied for their theoretical properties \cite{kulesza2012determinantal,gillenwater2014expectation}, and their machine learning applications including summarization \cite{lin2012learning,chao2015large}, object retrieval \cite{affandi2014learning} and conversation generation \cite{song2018towards}.
Structured Determinantal Point Processes (SDPPs) \cite{kulesza2010structured} extend the DPP model to handle an exponentially-sized set of particles (structures) via a natural factorization combined with the second-order message passing, which leads to tractable algorithms for inference.

Recent works use DPPs for item list recommendation in \cite{zhou2010solving,gartrell2016bayesian} to solve cross-item diversity, but to the best of our knowledge, there are no studies about cross-bundle diversity. In this paper, we use SDPP as the tool to decompose the quality/diversity term.

\section{Personalized Bundle List Recommendation}
\stitle{Notation.}
Let $\mathcal{I}$ be the whole set of items that one could select from and $|N|$ be the number of all items. A bundle $b$ is any combination out of $\mathcal{I}$ which consists of items, and $T$ denotes the number of items in a bundle. The items are unordered in bundles. When regarding the bundle as a sequence, we usually give the order by sorting the items of the bundle in terms of their prices in a descendant order, because the cheaper items are more substitutable for the bundle and could endure more exposure bias in the sequence generation model. Let $\mathcal{B}$ be the set of all possible bundles, so $|\mathcal{B}|$ equals $\sum_{t=1}^T \binom Nt$ with the fixed maximum bundle size $T$, where $\binom{N}{t} = \frac{N!}{t!(N-t)!}$ is the combination of $t$ out of $N$.
\begin{align*}
\mathcal{I} &= \{1, 2, \dots, N\} \qquad \qquad \qquad \quad \;\; |\mathcal{I}|=N \\
\mathcal{B} &= \{b=\{i_1, i_2, \dots, i_T\} | i \in \mathcal{I}\} \qquad |b|=T, \; |\mathcal{B}| = \sum_{t=1}^T \binom{N}{t} \\
\mathcal{Y} &= \{y=\{b_1, b_2, \dots, b_K\} | b \in \mathcal{B}\} \quad |y|=K, \; |\mathcal{Y}| = \binom{|\mathcal{B}|}{K}
\end{align*}

Here $K$ is the size of a recommended bundle list $y$. Still, we consider the order does not matter in the bundle list. Let $\mathcal{Y}$ be the set of all possible bundle list. The user context $C_u$ is usually represented as the item sequence he/she has clicked/bought in the history.

\stitle{Problem Formalization.}
Let $\tilde{\mathcal{P}}(\cdot, \cdot): \mathcal{Y} \times \{C_u\} \rightarrow \mathbb{R}^+$ be a universal compatibility function measuring the score (compatibility) of the bundle list $y$ and the user context $C_u$, and we refer $\tilde{\mathcal{P}}(y|C_u)$ to a unnormalized conditional probability of $y$ given $C_u$. Then, the personalized bundle list recommendation problem can be formalized as a structured prediction problem \cite{Belanger2015Structured}, which tries to find the bundle list $\hat{y}$ satisfying:
\begin{equation}
\hat{y} = \arg \max_{y \in \mathcal{Y}} \tilde{\mathcal{P}}(y=\{b_1, b_2, \dots, b_K\}|C_u)
\end{equation}

\eat{There are many ways to define a reasonable compatibility function, e.g., by a weighted linear combination or product of some precision metrics and diversity metrics.}

Let $\mathcal{P}(y|C_u)$ be the normalized conditional probability of $\tilde{\mathcal{P}}$ over $y$. $\mathcal{P}(y|C_u)$ is actually modeling a distribution over the \emph{doubly exponential} number of $N$, so there are $\binom{\sum_{t=1}^T \binom{N}{t}}{K}$ possible structures of $y$ given $C_u$ naively, which poses an extreme computational challenge. In this paper, we consider the compatibility function $\tilde{\mathcal{P}}$ related to both the quality and diversity.

The bundle list recommendation is a generalization of the single bundle recommendation problem and single-item list recommendation problem. When each bundle is independent given the user context, we have $\mathcal{P}(y=\{b_1, b_2, \dots, b_K\}|C_u)) = \prod_k \mathcal{P}(b_k|C_u)$. In this case, there are no correlations among the bundle list, and the problem becomes the single bundle recommendation problem \cite{zhu2014bundle}, sometimes considered as the personalized bundle ranking problem \cite{Pathak2017Generating}. Furthermore, when the bundle size $T$ equals one it is the single-item list recommendation problem.

\eat{There are several main challenges to this problem. First, We could notice that $y$ is actually modeling a structure over the \emph{doubly exponential} number of the number of items $N$. If there are $N^t$ possible bundles with the fixed bundle size $t$, there are $2^{N^t}$ possible structures for $y$, which poses an extreme computational challenge. Second, Compared with the single-item list recommendation task, diversity is a much more critical metrics in bundle recommendation task \cite{Pathak2017Generating} since bundles with very similar items may reach a high ranking scores at the same time.}

\eat{As we notice that unlike the traditional item recommendation, it is intractable to produce the bundle list by applying the ranking method directly in all possible bundles, so we need a method to generate bundles efficiently for each user.}

\section{Bundle Generation Network}

Here we propose a \emph{Bundle Generation Network} (BGN) to solve the personalized bundle list recommendation problem considering both the quality and diversity.

In the following subsections, we introduce how to decompose the bundle list probability into the quality/diversity part by SDPP and factorize them respectively. Then, we illustrate the measurement of diversity and design an improved neural network with the feature-aware softmax to learn the quality part. Besides, we propose a masked beam search to control the bundle size and avoid duplicated items. Finally, we demonstrate how to integrate these parts to BGN to tackle the process of generation for the bundle recommendation.

\subsection{SDPP-based Bundle Generation Model}
We define a \emph{structured determinantal point process} (SDPP) \cite{kulesza2010structured} on the probability space $(\mathcal{B}, \mathcal{Y}, \mathcal{P})$. When $\mathbf{Y}$ is a random variable of bundle recommendation list drawn according to the probability mass function (PMF) $\mathcal{P}$, the distribution is given by:
\begin{equation}
\mathcal{P}(\mathbf{Y}=y) \triangleq det(L_y) / Z \label{eqa: dpp}
\end{equation}

Here, $L$ is a symmetric positive semi-definite kernel of SDPP, with the size of $|\mathcal{B}|*|\mathcal{B}|$. $L_y$ denotes the restriction of $L$ to the entries indexed by elements of $y$, and $Z=\sum_{y' \in \mathcal{Y}}det(L_{y'})$ is the normalization constant. There is an intuitive geometric interpretation of the kernel $L$, that is, $\det(L_y)$ is the square of the volume spanned by its associated feature vectors \cite{kulesza2012determinantal}. Notice that, each element of $y$ refers to a bundle which has a certain structure, so we consider that the DPP is structured.

One of the benefits of defining a DPP probability model is that we can decompose quality and diversity in a very natural way. For any element $L_{ab}$ of positive semi-definite matrix $L$ we could always find a quality/diversity decomposition:
\begin{equation}\begin{split} \label{eqa: quality/diversity decomposition}
L_{ab}
&= q(a) \phi(a)^T \phi(b) q(b) \\
&= q(a) S(a, b) q(b)
\end{split} \end{equation}
where $a, b \in \mathcal{B}$, $L_{ab}$ denotes the element indexed by bundle $a$ and $b$, $q(b) \in \mathbb{R}$ denotes the quality term of bundle, and $\phi(b) \in \mathbb{R}^D$ denotes the normalized diversity feature with $||\phi(b)||_2=1$. Here we define a similarity function/metric $S(a, b): \mathcal{B} \times \mathcal{B} \rightarrow \mathbb{R}$ which satisfies $S(a, b) = \phi(a)^T \phi(b)$ and keeps the corresponding similarity matrix $S$ positive semi-definite. We reformalize Equation (\ref{eqa: quality/diversity decomposition}) in the matrix way as \begin{equation}
L=QSQ
\end{equation}
where $Q$ denotes a diagonal matrix $\text{diag}\{q(b)_{b \in \mathcal{B}}\}$. The decomposition balances the quality and the diversity.
Then, the optimization criterion of the personalized bundle list recommendation is to find a bundle list $y$ maximizing the $\mathcal{P}(y|C_u;\theta)$ given the user context $C_u$:
\begin{gather}
\eat{\arg \max_{y} \; \mathcal{P}(L_y|C_u;\theta) \nonumber \\
\Leftrightarrow} \arg \max_{y} \; \det(S_y) \cdot \prod_{b \in y} q^2(b|C_u;\theta) \label{eqa: maximize_DPP}
\end{gather}

\eat{composed by the similarity function $S(a, b)$. The decomposition balances the quality term $q(b)$ and the similarity function $S(a, b)$. \eat{We consider the similarity function is independent of user context $C_u$ and could be constructed by domain knowledge in advantage.} Note that actually we do not need to compute and store the whole matrix $S$ but only need to pre-define the similarity function $S(a, b)$ because here we only need to compute $L_y$ rather than $L$ as we shall introduce later.}

One major challenge for solving the personalized bundle list recommendation is that, the optimization criterion of the bundle list has the complexity over doubly exponential number of $N$. According to the decomposition in Equation (\ref{eqa: maximize_DPP}), we reduce the complexity to a quality term $q$ over the measure space $\mathcal{B}$, and a diversity term $S$ over the measure space $\mathcal{B} \times \mathcal{B}$. However, both of the terms still have exponential space which is intractable.

SDPP considers that any elements in measure space $\mathcal{B}$ can factorize over a set of factors $F$ \cite{kulesza2010structured}, where a factor $\alpha \in F$ is a small subset of the parts of a structure. Then, we continue to factorize the quality term and the diversity term respectively:
\begin{equation}
q(b) = \prod_{\alpha \in F} q_\alpha(b_\alpha) \qquad
\phi(b) = \sum_{\alpha \in F} \phi_\alpha(b_\alpha \label{eqa: phi_decompose})
\end{equation}

The factorization defines how we interpret the model structure and keeps the model tractable. Then we shall specify the factorization for the diversity and quality respectively, and integrate them to produce high-quality and diversified bundle list.

\subsection{The Measurement and Factorization of Diversity}

The diversity we measure in our work is a interpretable metric/goal, trying to quantify the user-experience in bundle recommendation system, which is independent of the likelihood from data. This implicit relationship learned from data may not be the same `diversity' as we target, and the correlations among items from data have been considered into the quality part. Whereas, the explicit similarities defined by side information and prior knowledge provide exact signals for the purpose of the measurement and adjustment.\eat{ The decomposition gives a way to combine prior knowledge and the neural network.} Besides, the recommendation system is usually lake of the diversity-labeled ground-truth list, which makes it hard to supervise the complete likelihood.

With Equation (\ref{eqa: quality/diversity decomposition}) and Equation (\ref{eqa: phi_decompose}) and , we factorize the bundle similarity function to the item similarity function:
\begin{align}
S(a, b)
&= \phi(a)^T \phi(b) \nonumber \\
&= (\sum_{\alpha \in F}\phi_\alpha(a_\alpha)^T) (\sum_{\beta \in F}\phi_\beta(b_\beta)) \nonumber \\
&= \sum_{\alpha, \beta} \phi_\alpha(a_\alpha)^T \phi_\beta(b_\beta) \nonumber \\
&= \sum_{i \in a, j \in b} s_{a, b}(i, j)
\end{align}

Here, $s_{a, b}(i, j)$ measures the similarity of items rather than bundles, which is independent of user context. The similarity funciton usually should be consistent with the metrics of diversity predefined according to domain knowledge. For example, in our experiments the $s_{a, b}(i, j)$ is given by Jaccard similarity of bundle:
\begin{equation}
s_{a, b}(i, j) = \frac{1}{|a \cup b|} \delta(i=j) \label{eqa: straightforward_sim_metric}
\end{equation}
where $\delta(\cdot)$ denotes the indicator function. Note that the bundles do not contain duplicated items inside. This function can maintain the positive semi-definition of $L$ \cite{bouchard2013proof} and prevent the identical items shown in the different bundles. The function keeps consistent local measurement with Equation (\ref{eqa: metric_div}), and SDPP optimizes the global negative correlations. Besides, the metric of diversity could also be defined by calculating the similarity of attributes\eat{ or the overlap of category hierarchy of items}, which provides the fine-grained diversity of attributes. \eat{However, some inappropriate function definitions may break the semi-definiteness of L, it is not a major concern in practice because we could let $L=L+ \epsilon I$.}

\subsection{Modeling the Quality Part with Feature-Aware Softmax}

Here, we focus on modeling the quality part $q(b|C_u; \theta)$ by a typical \emph{Encoder-Decoder} framework, where the encoder extracts the information from the user context and the decoder generates the high-quality candidate bundle list with a proposed feature-aware softmax.

For the encoder, the user context can be represented as the purchase history of items. We use a text-CNN \cite{kim2014convolutional} to extract the information, which has recently been shown of effectiveness in text classification and information extraction \cite{cnn_wsdm,atrank}. We show the basic network architecture in Figure \ref{fig:context_bundle_net} and omit the discussion of the details because it is not one of our main contributions. The text-CNN is parallelizable and faster than bi-LSTM, and it could also be easily improved by Transformer \cite{vaswani2017attention} based on the self-attention architecture in the future work.

For the decoder, the quality of a bundle $q(b|C_u; \theta)$ is an unnormalized joint probability parameterized in terms of a generic $\theta$, so we factorize it in the Bayesian formula way. Note that the Bayesian formula always holds and models the complete distribution of quality, the assumption here is sharing the parameters in a sequence way.
\begin{align}
q(b | C_u; \theta)
\propto \;& p(b = \{i_1, i_2, \dots, i_T\} | C_u; \theta)  \nonumber \\
=\;& \prod_{t=1}^{T} p(i_t | i_1, ..., i_{t-1}, C_u; \theta) \label{eqa: bayesian_decompose}
\end{align}

In Equation (\ref{eqa: bayesian_decompose}), we factorize the joint probabilities of any bundle by multiplying all conditional probabilities for each of its elements. By sharing the same weights of those conditional probabilities, we use a sequence generation model with the (stacked) LSTM cell with the well-known Luong Attention \cite{luong2015effective} to calculate the probability. Let $x_t$ be the decoder's input and $t$ be the item position in the bundle. We initialize $h_0$ with the encoder's output. Still, we omit the common details of LSTM and attention.
\begin{gather}
h_0 = \text{text-CNN}(C_u) \nonumber \\
x_{t} = [\text{item\_emb}(i_t), \text{cate\_emb}(i_t), ...] \nonumber \\
h_{t} = \text{stacked-LSTM-attention}(h_{t-1}, x_{t-1}, C_u) \nonumber \\
p(i_t|i_1, ..., i_{t-1}, C_u; \theta) = \text{feature-aware-softmax}(h_{t})_{i_t}
\end{gather}

The traditional softmax layer contains a weight matrix $W^{(h+1)*N} = [w_1, w_2, ..., w_N]$, where $w_i$ could be considered as an embedding of item\_id of $i$. Note that, we eliminate the bias term here just for writing convenience.
\begin{equation}
\text{softmax}(h_t, W)_j = \frac{exp(h_t^T w_j)}{\sum_{\hat{j}=1}^N exp(h_t^T w_{\hat{j}})}
\end{equation}

However, the item candidates in recommendation system usually have more belonging features rather than just item\_id. The traditional softmax could result in inadequate representation for weight matrix owing to the lack of features.

Inspired by pointer network \cite{vinyals2015pointer}, we propose \textbf{feature-aware (FA) softmax}\label{feature aware softmax}, which modifies the softmax operator with feature-aware weight matrix instead of a stationary weight matrix, and makes the softmax sensitive to dynamic information. Feature-aware softmax uses the dynamic weight matrix built by $F(x_i)$, where $F$ is a nonlinear transformation function for all kinds of features, so that the softmax could be sensitive to features besides item\_id.
\begin{gather}
E^{(h+1)*N} = [e_1, e_2, ..., e_N]^1 , \qquad e_i = F(x_i) \nonumber \\
\text{feature-aware-softmax}(h_t)_j = \text{softmax}(h_t, E)_j \label{eqa: feature_aware_softmax}
\end{gather}

\eat{We argue the effectiveness of feature-aware softmax again by proving that the sampled feature-aware softmax cross entropy loss is equivalent to optimizing average BPR at each step under certain condition.}

We use the co-purchased data $\mathcal{D}=\{b,C_u\}_{1:|\mathcal{D}|}$ to guide the learning process. The loss could be defined as the mean cross entropy for the softmax layer of each step, which guarantees the same form as Equation (\ref{eqa: bayesian_decompose}) and the model is to maximize the likelihood of a single bundle.
\begin{equation}
\mathcal{L}_{\text{MLE}} = - \frac{1}{T} \sum_{t=1}^T \log p(i_t|i_1, ..., i_{t-1}, C_u; \theta)
\end{equation}

However, the feature-aware softmax might be time-consuming during training owing to constructing the weight matrix $E$ dynamically. So we apply feature-aware softmax with sampled softmax loss \cite{jean2014using}, denoted as the sampled feature-aware softmax loss, which accelerates the training process.

We show that the sampled feature-aware softmax is a generalization of optimizing the average BPR \cite{bpr} at each step. The sampled feature-aware softmax samples a part of negative items at each step plus the ground truth of the next step, and then builds the corresponding weight matrix. We regard $h_t$ as the user vector of BPR, $e_{i_t}$ as the positive item vector, and $e_{s_t}$ as the negative item vector, where $s_t$ is the sampled negative item at step $t$. When setting the sample number to be 1, the loss of the sampled feature-aware softmax is given by:
\begin{align}
\mathcal{L}_{\text{sample-1}}
&= - \frac{1}{T} \sum_{t=1}^T \log
\frac{exp(h_t^T e_{i_t})}
{exp(h_t^T e_{i_t}) + exp(h_t^T e_{s_t})}
+ \lambda_{\theta} || \theta ||^2 \nonumber \\
&= - \frac{1}{T} \sum_{t=1}^T \log
\sigma (h_t^T (e_{i_t} - e_{s_t}))
+ \lambda_{\theta} || \theta ||^2 \nonumber \\
&= - \frac{1}{T} \sum_{t=1}^T \log p(\theta | \ge_{t, C_u}) = \frac{1}{T} \sum_{t=1}^T \mathcal{L}_{\text{BPR}}(t) \label{eqa: sample_fa_softmax}
\end{align}
where $\sigma(\cdot)$ is the logistic sigmoid function, and $\ge_{t, C_u}$ is the pairwise preference at each step as \cite{bpr}. So the feature-aware softmax with one negative item is equivalent to optimizing average BPR at each time in equation (\ref{eqa: sample_fa_softmax}).

\eat{One drawback of our model is that we maximize the likelihood of $p(b)$ instead of $p(y)$ during learning, which leads to the learning process not taking into account the diversity term among items. On the one hand, it is similar to the difficulty encountered by many list-wise methods, which is hard to construct the real probability of $p(y)$ from groundtruth data. On the other hand, we observe that the straightforward way works well in practice.}

\subsection{Overall Process of Bundle Generation}

The goal of bundle list recommendation is to generate the high-quality and diversified bundles list $y$ maximizing the $p(y|C_u)$ in Equation (\ref{eqa: maximize_DPP}). We integrate and summarize the overall process of bundle generation in this subsection.

\begin{figure*}[htbp]
	\centering
	\includegraphics[scale=0.55]{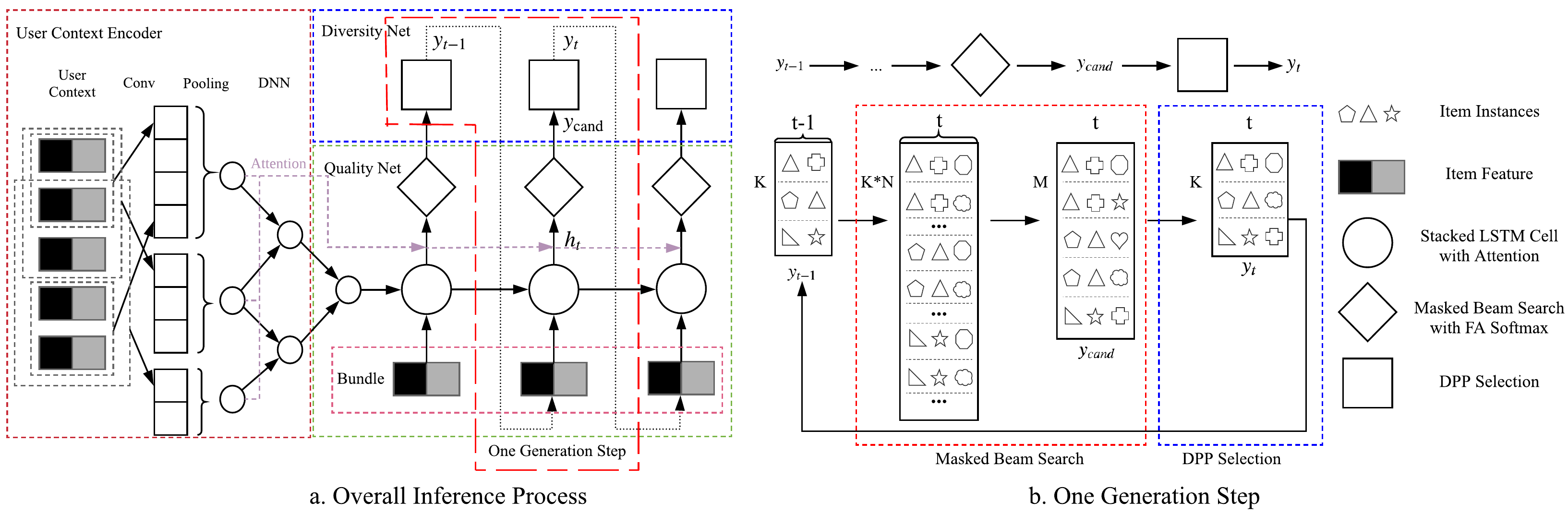}
	\vspace{-0.6cm}
	\caption{Inference Process of Bundle Generation Network.}
	\label{fig:context_bundle_net}
	\vspace{-0.4cm}
\end{figure*}

Figure \ref{fig:context_bundle_net}(a) shows the overall inference process of BGN. We initialize $h_0$ with the output of text-CNN and then produce the bundle list by expanding the generated prefix of bundles at each step.
Figure \ref{fig:context_bundle_net}(b) shows how the bundle list $y_t$ expands at each generation step. Each small shape icon denotes an item, each row of the rectangular block is a generated bundle prefix, and $t$ indicates the aligned size of the generated bundle prefix. At each generation step, we use the beam search to produce $K*N$ candidate bundles. Beam search \cite{wiseman2016sequence} is a heuristic search algorithm that explores the search space by expanding the most promising item in a limited set, which is widely used  \cite{seq2seq} in sequence generation approaches. Here, we use the beam search to prune the low-quality ones down to $y_{\text{cand}}$ with width $M$, which is crucial for the efficiency. The submodular assumption \cite{kulesza2012determinantal} of DPPs is widely used during inference in practice including recommendation systems \cite{gartrell2016bayesian,fast_greedy_dpp}, where one can find the solution in polynomial time. Following Equation (\ref{eqa: maximize_DPP}), we choose the bundle $b$ from $y_{\text{cand}}$ one by one maximizing $P(y \cup \{b\})$, which is given by:
\begin{align}
\eat{& \arg \max_{b \in y_{\text{cand}}} \; \log P(\hat{y} \cup \{b\}) \nonumber \\
	\Leftrightarrow} & \arg \max_{b \in y_{\text{cand}}} \; \log \det S_{y \cup \{b\}} \nonumber + \sum_{\beta \in y \cup \{b\}} \log q^2(\beta | C_u) \\ \eat{
	\Leftrightarrow & \arg \max_{b \in y_{\text{cand}}} \; \log \det S_{y \cup \{b\}} + 2 * \log q(b | C_u) \nonumber \\}
	\Leftrightarrow & \arg \max_{b \in y_{\text{cand}}} \; \log p(b | C_u) + \lambda \log \det S_{y \cup \{b\}} \label{eqa: greedy_map}
\end{align}

Here, $\lambda$ is a hyper-parameter controlling the tradeoff between quality and diversity. When we increase $\lambda$, we pay more attention to the diversity among the generated bundles than the precision of the final list, and vice versa.

In addition, we need to avoid duplicated items in the bundle, and would like to generate bundles with the larger size in practice. Usually, We add a particular end token of an item to mark the end of bundle generation so that we can produce the various-length bundle. However, BGN uses the co-purchase data or pre-defined bundles to supervise the quality part. The distribution of groundtruth bundle size may not be consistent with expected distribution for target applications. In practice, one may need larger bundle size due to sellers' promotion. Besides, the traditional beam search often has a significant length bias and squeezes the distribution of bundle size to a much more skewed one that prefers shorter size, as we show in Figure \ref{fig:bs0} in our experiments. If we simply select the dataset with larger bundle size when training, it could cause more sparsity issue owing to the reduction of the training set.

For controlling the distribution of bundle size, as well as eliminating duplicated items, we improve the softmax with masks during beam search, which denotes the masked beam search. The masked beam search subtracts a mask vector $m_t \in \mathbb{R}^N$ to the logits for decreasing the probability in beam search during inference, which is given by:
\begin{equation}
\text{masked\_softmax}(h_t, E, m_t)_j = \frac{exp(h_t^T e_j - m_{t, j})}{\sum_{\hat{j}=1}^N exp(h_t^T e_{\hat{j}} - m_{t, \hat{j}})}
\end{equation}
where $m_{t, j}$ is the $j$-th element of $m_t$. Notice that, the mask vector reduces the certain unnormalized probabilities exponentially. In the bundle recommendation problem, we set $m_t$ to be:
\begin{equation}
m_{t,j} = \begin{cases}
\max (C - t, 0) \qquad & \text{j is the end token} \\
+\infty \qquad & \text{j has been generated before $t$} \\
0 \qquad & \text{otherwise}
\end{cases} \label{eqa: shifting_parameter}
\end{equation}
where $C$ is a positive hyper-parameter for shifting the distribution of the bundle size, which gives a threshold for end token. The reduction effect due to the shifting hyper-parameter $C$ decays exponentially with time step and dribbles away after $C$ steps.\eat{We show the impact of the shifting parameter $C$ in Figure \ref{fig:masked_beam_search_distribution} in our experiments.} Besides, the mask vector can avoid the duplicated elements in the bundle. We can set $m_{t,j}$ to be infinity when we definitely don't want to generate the item $j$ at time step $t$. \eat{For example, we can use this trick to produce a bundle where all items belong to only one certain shop in a practical application.}

\eat{BGN uses the co-purchase data or pre-defined bundles to supervise the quality part. However, the sequence generation methods usually suffer from generating short sequences \cite{koehn2017six}. Besides, in practice, we may need larger bundle size due to sellers' promotion. So we propose the \textbf{masked beam search} to control the size of the generated bundles.

However, the traditional beam search tends to generate short sequences \cite{koehn2017six}. }

The overall algorithm is shown in Algorithm \ref{algo: greedy}, where line 5-15 show the single generation step and line 10-12 show the DPP selection according to Equation (\ref{eqa: greedy_map}).
Each generation step consists of the stacked LSTM cell, the masked beam search with feature-aware softmax, and the DPP selection. We consider the hidden dimension of LSTM cell is constant here. The time complexity of generating candidate list $y_{\text{cand}}$ using beam search is $O(TMN)$, and we usually choose $M$ larger than $K$ to ensure that the candidate set is sufficiently diversified. The time complexity of the DPP selection at each step is $O(MK^\omega)$ where the $K^\omega$ term comes from the determinant calculation and $\omega$ denotes $2.37$. Note that we do not need $O(N^2)$ space to compute and store all item pair's similarity because we only compute $S_y$ rather than entire $S$. The total time complexity of inference process for BGN is $O(TMN+TMK^\omega)$.

\begin{algorithm}[htb]
	\caption{Bundle Generation Network during Inference}
	\label{algo: greedy}
	\begin{algorithmic}[1]
		
		\State \textbf{Input:} User context $C_u$, Similarity function $S(a, b)$, Quality/diversity hyper-parameter $\lambda$, Maximum bundle size $T$, Beam search width $M$, Output list size $K$
		\State \textbf{Output:} Diversified bundle list $y$ with size $K$
		
		\State $h_0 \leftarrow \text{text-CNN}(C_u)$
		\State $y_0 \leftarrow \varnothing$
		\State Building the weight matrix $E$ of the feature-aware softmax according to Equation (\ref{eqa: feature_aware_softmax})
		\For{$t \in [1, ..., T]$}
		\State $h_t \leftarrow$ Stacked LSTM Cell with Attention$(h_{t-1}, y_{t-1}, C_u)$
		\State $y_{\text{cand}} \leftarrow$ Masked Beam Search with FA Softmax$(h_t, E, m_t)$
		\State $\text{//} \: y_{\text{cand}}$ contains $M$ bundles with maximum size $t$
		\State $y_t \leftarrow \varnothing$
		
		\State $\text{//}$ DPP Selection
		\While{$|y_t| \neq K$}
		\eat{\State $b \leftarrow \arg \max_{b \in y_{\text{cand}}} \log p(b| C_u) + \lambda \log \det S_{y_t \cup \{b\}}$}
		\State Choose $b$ from $y_{\text{cand}}$ according to Equation (\ref{eqa: greedy_map})
		\State $y_{\text{cand}} \leftarrow y_{\text{cand}} - \{b\}$
		\State $y_t \leftarrow y_t \cup \{b\}$
		\EndWhile
		
		\State $\text{//} y_t$ has K bundles with maximum bundle size $t$
		\EndFor
		\State \Return $y \leftarrow y_T$
		
	\end{algorithmic}
\end{algorithm}

\section{Experiment}
We organize the experiment section as follows. First, we clarify the experiment setup, measurement metrics, and all competitors. Second, we illustrate the performance gain of feature-aware softmax and the performance impact of the masked beam search. Then, we show the balances the diversity and the precision. Finally, we compare with all competitors to see the advantages of our methods on metrics of precision, diversity, response time for generation and AUC for ranking.

\subsection{Experiment Setup}

\stitle{Dataset.}
We use three public datasets and one industrial dataset: Electro and Clothe are two public datasets of Amazon in \cite{amz_mc} where bundles are generated from co-purchase records synthetically. Steam is a public dataset where bundles are extracted from video game distribution network in \cite{Pathak2017Generating}. Taobao is an industrial dataset where we collect the real sales transaction of bundles from the online bundle list recommendation services. We show the statistics for each dataset in Table \ref{table: datasets}.

\begin{table}[htbp] 
\begin{center}
	\begin{tabular}{|c|cccc|}
		\hline
		Dataset & Electro & Clothe & Steam & Taobao \\ \hline
		Items & 63,001 & 23,033 & 10,978 & 104,328 \\
		Users & 192,403 & 39,387 & 88,310 & 2,767 \\
		Categories & 801 & 484 & - & 3,607 \\
		Records & 1,689,188 & 278,677 & 87,565 & 54,344 \\
		Bundles & 327,874 & 72,784 & 615 & 1,236 \\
		Average Bundle Size & 1.50 & 2.47 & 5.73 & 4.06 \\ \hline
	\end{tabular}
	\vspace{0.2cm}
	\caption {Statistics of each dataset.}
	\label{table: datasets}
	\vspace{-0.8cm}  
\end{center}
\end{table}

Amazon Dataset\footnote{http://jmcauley.ucsd.edu/data/amazon}. We use two subsets (Electro and Clothe) of Amazon product data in \cite{amz_mc}, which have already been reduced to satisfy the 5-core property, such that each of the remaining users and items has five reviews. We regard the co-purchase items as a bundle, where items are sorted in terms of their prices in a descendant order. The behavior history of each user can be represented as $(b_{1}$, $b_{2}$, ..., $b_{k}$, ..., $b_{K})$ where $b_{k}$ is a bundle consisting of co-purchase items. We make the first $k$ bundle behaviors of each user as the user context to predict the $k+1$ th bundle behavior in the training set and validation set, where $k=1,2, ..., K-2$, and we use the first $K-1$ bundle behaviors to predict the last one in the test set. The average bundle size is 1.50 for Electro and is 2.47 for Clothe, and we notice that many bundles consist of only one item in these two datasets. For alleviating the sparsity of bundle purchase records and keeping enough training records, we consider the bundles with one item is valid in training set. However, in the test set, we only use bundles consisting of more than one items to measure the performance, which isolates the measurement from the single-item list recommendation. The features that we use contain user id, item id, category id, and bundle id. The category is a feature of items.

Steam Dataset\footnote{http://jmcauley.ucsd.edu/data/steam.tar.gz}. This dataset is collected in \cite{Pathak2017Generating} from the Steam video game distribution platform. The item is a certain game in this dataset. We partition all purchase records into 70\%/10\%/20\% training/validation/test splits. As in Amazon dataset and Taobao dataset, Some combinations of items in the test set may never be seen in the training set, which leads to the difficulty in generation. Note that there is no feature of the category in the Steam dataset.

Taobao Dataset. This dataset is collected from an online bundle list recommendation service of a commercial e-business, Taobao\footnote{https://www.taobao.com}. Different from the synthetic bundles consisting of co-purchase items in Amazon dataset, the bundles in the Taobao dataset are pre-defined manually by the sellers. Correspondingly, they have higher quality and longer average bundle size. We conduct the careful collection which guarantees that each user has at least three bundle behaviors. We use the last one of bundle behaviors in the test set and the others in the training set. The user context consists of items which users have purchased. The features that we use contain user id, item id, category id, and bundle id.

\stitle{Measurements.} 
For the bundle generation problem, the metrics we mainly consider are the top-k precision $\textbf{\textit{pre@k}}$ measuring the quality and the cross-bundle diversity $\textbf{\textit{div}}$. The $\textit{pre@k}$ is the average precision with the ground truth over each user and each position in the final bundle list. The $\textit{div}$ is defined as $1-\text{Jaccard Similarity}$ over each pair of the final bundle list.
\begin{gather}
pre@k = \frac{1}{|U|} \sum_u \frac{1}{|K|} \sum_{b \in y_u} \; \frac{|b \cap gt_u|}{|b \cup gt_u|} \\
div =\frac{1}{|U|} \sum_u \frac{1}{|K|(|K|-1)} \sum_{a, b \in y_u}{( 1 - \frac{|a \cap b|}{|a \cup b|} )} \label{eqa: metric_div}
\end{gather}
where $gt_u$ is the ground truth bundle that a user buys or clicks. According to the dataset setup above, there is only one bundle in $gt_u$ for the Amazon and Taobao dataset, and there are multiple bundles in $gt_u$ for the Steam dataset. Let $y_u$ be the final recommendation bundle list with size $K$, $U$ be all users in the test set. These two metrics can be applied to any case no matter whether the bundle is predefined. Note that, the repeated items belonging to different bundles hit the ground truth and increase the $\textit{pre@k}$, but they decrease $\textit{div}$. 

For studying whether our method can adapt to the bundle ranking problem, we use the \textbf{AUC} to evaluate the performance, which is given by:
\begin{equation}
	AUC=\frac{1}{|U|} \sum_{u} \frac{1}{|gt_u|} \sum_{b^+ \in gt_u} \delta(\hat{p_{u,b^+}} > \hat{p_{u,b^-}})
\end{equation}
where $\hat{p_{u,b}}$ is the predicted probability that a user may act on a bundle and $\delta(\cdot)$ denotes an indicator function. Note that, $b^-$ is sampled uniformly from all valid bundles in the dataset, which are all co-purchase records consisting of more than two items for the Amazon dataset and all pre-defined bundles for the Steam and Taobao dataset.

\stitle{Competitors.} We mainly consider six methods as below:
\begin{itemize}
	
	\item \textbf{BGN}. This denotes our Bundle Generation Network method. For CNN encoder of user context, we use various filters with the window size of $\{1, 2, 4, 8, 12, 16, 32, 64\}$, and each filter contains 12 output channels, so the final dimension of output channel is 96. The other latent dimension is set to be 64. We use two layers' stacked LSTM cell with attention in the decoder. The batch size is set to be 16. We use l2-loss in the training process, and the weight is set to be 5e-5. Adam is adopted as the optimizer. The beam search width is set to be 50. The sample number of sampled feature-aware softmax is set to be 1024. We use Equation (\ref{eqa: straightforward_sim_metric}) as the similarity function. $\lambda$ is the quality/diversity hyper-parameter ranging from $0.0$ to $5.0$ with step $0.25$. $C$ is the shifting hyper-parameter controlling the distribution of bundle size, which ranges from $0$ to $20$ with step $1.0$. \eat{The $\lambda$ and $C$ are set to be $0$ in default under the consideration of controlling variates.}
	
	\item \textbf{Freq}. This is a well-known frequent itemsets mining algorithm introduced in \cite{apriori}, and we generate bundles according to frequent itemsets discovered in co-purchase or pre-defined bundles. Then we recommend the most frequent $k$ bundles to users. Note that this is not a personalized method.
	
	\item \textbf{BRP}. \emph{BRP} (Bundle Recommendation Problem) \cite{zhu2014bundle} is a personalized single bundle recommendation algorithm, which optimizes the expected reward based on Mixed-Integer Quadratic Programming (MIQP). We use matrix factorization as the probability model which has the best performance in the original paper, and use Gurobi\footnote{http://www.gurobi.com/}, one of the fastest public mathematical programming solver, to optimize the MIQP problem.
	
	\item \textbf{BBPR}. \emph{BBPR} (Bundle Bayesian Personalized Ranking) is proposed by \cite{Pathak2017Generating}. It builds the optimization criterion upon the Bayesian Personalized Ranking(BPR) \cite{bpr}, with the mean pair-wise Pearson correlation of the items. Combined with traditional matrix factorization techniques, it utilizes both item and bundle information including bundle size and item compatibility for better performance. As for the bundle generation, it proposes a greedy algorithm following an annealing schedule to give personalized bundles, which avoids ranking for all possible bundles. In the Steam dataset, we maintain the setup of typer-parameters as the original paper, and in other datasets, we deploy a small range of fine-tuning and report the best results.
	
	\item \textbf{RankAll}. \emph{RankAll} ranks all the existing bundles in the training set and returns the top-K results, and we consider it as a strong baseline to validate whether the ranking methods are good enough for the bundle generation task regardless of the consumption of time and space. Note that, \emph{RankAll} refers to rank all the existing bundles in the training set, instead of ranking all possible bundles in $\mathcal{B}$, which is too large to enumerate even for modern machines. Specifically, \emph{RankAll} adopts the same loss as \emph{BBPR}, but utilizes two text-CNN networks to extract the user context and the bundle information respectively. The hyper-parameter setting of text-CNN is agree with \emph{BGN}. Ranking all the existing bundles in the training set is time-consuming and usually is unacceptable in practice.
	
\end{itemize}

\stitle{Environment.} All experiments are conducted on a machine with one GPU (NVIDIA Tesla P100 16G) and 96 CPUs (Intel Xeon Platinum 2.50GHz). We use Tensorflow \footnote{https://tensorflow.org} as the main toolkit of neural network.

\subsection{Feature-Aware Softmax Performance Gain}

The feature-aware softmax combines more features besides the item\_id in the weight matrix of softmax. We demonstrate the performance gain of all datasets in Table \ref{seq2seq improve}, where `w/o FA' denotes `without feature-aware' and `w/ FA' denotes `with feature-aware' for short.

\begin{table}[htbp] 
	\centering
	\begin{tabular}{|c|c|c|c|c|c|}
		\hline
		\multicolumn{2}{|c|}{Dataset} & Electro & Clothe & Steam & Taobao \\ \hline
		\multirow{2}{*}{pre@10} & BGN w/o FA & 0.32\% & 0.13\% & - & 1.75\% \\
		& BGN w/ FA & 0.60\% & 0.38\% & 9.65\% & 2.82\% \\ \hline
		\multirow{2}{*}{pre@5} & BGN wo/ FA & 0.38\% & 0.12\% & - & 1.83\% \\
		& BGN w/ FA & 0.67\% & 0.43\% & 14.96\% & 3.09\% \\ \hline
	\end{tabular}
	\vspace{0.2cm}
	\caption{Feature-Aware (FA) softmax Performance Gain}
	\label{seq2seq improve}
	\vspace{-0.8cm}
\end{table}

In Table \ref{seq2seq improve}, we observe that the proposed feature-aware softmax can improve the precision significantly, which yields the improvements on $pre@10$ by 0.87x, 1.92x and 0.61x on Electro, Clothe and Taobao dataset respectively. The gain brought by feature-aware softmax indicates that, compared to the traditional softmax, the feature-aware softmax considers more features in the weight matrix and alleviates the sparsity issue of item\_id effectively in bundle generation. The performance gain also verifies the importance of features for recommendation tasks as in \cite{wide_n_deep}. Note that there is no feature of categories in the Steam dataset, so there is no difference with feature-aware softmax.

Note that all competitors except \emph{Freq} consider the same complete features as BGN, but BGN with feature-aware softmax performs better because we factorize the bundle probability as a sequence of conditional probability, which is equivalent to optimizing generalized BPR loss directly at each step as shown in Equation (\ref{eqa: sample_fa_softmax}).

\subsection{Performance Impact due to controlling the Bundle Size}

The distribution of the training data may not be consistent with the distribution of the bundle size for the target application. According to the masked beam search, we study the performance impact due to controlling the Bundle Size.

\begin{figure}[htbp]
	\begin{minipage}[t]{0.495\linewidth}
		\centering
		\subfigure[Hyper-parameter C]{
			\includegraphics[width=1.65in]{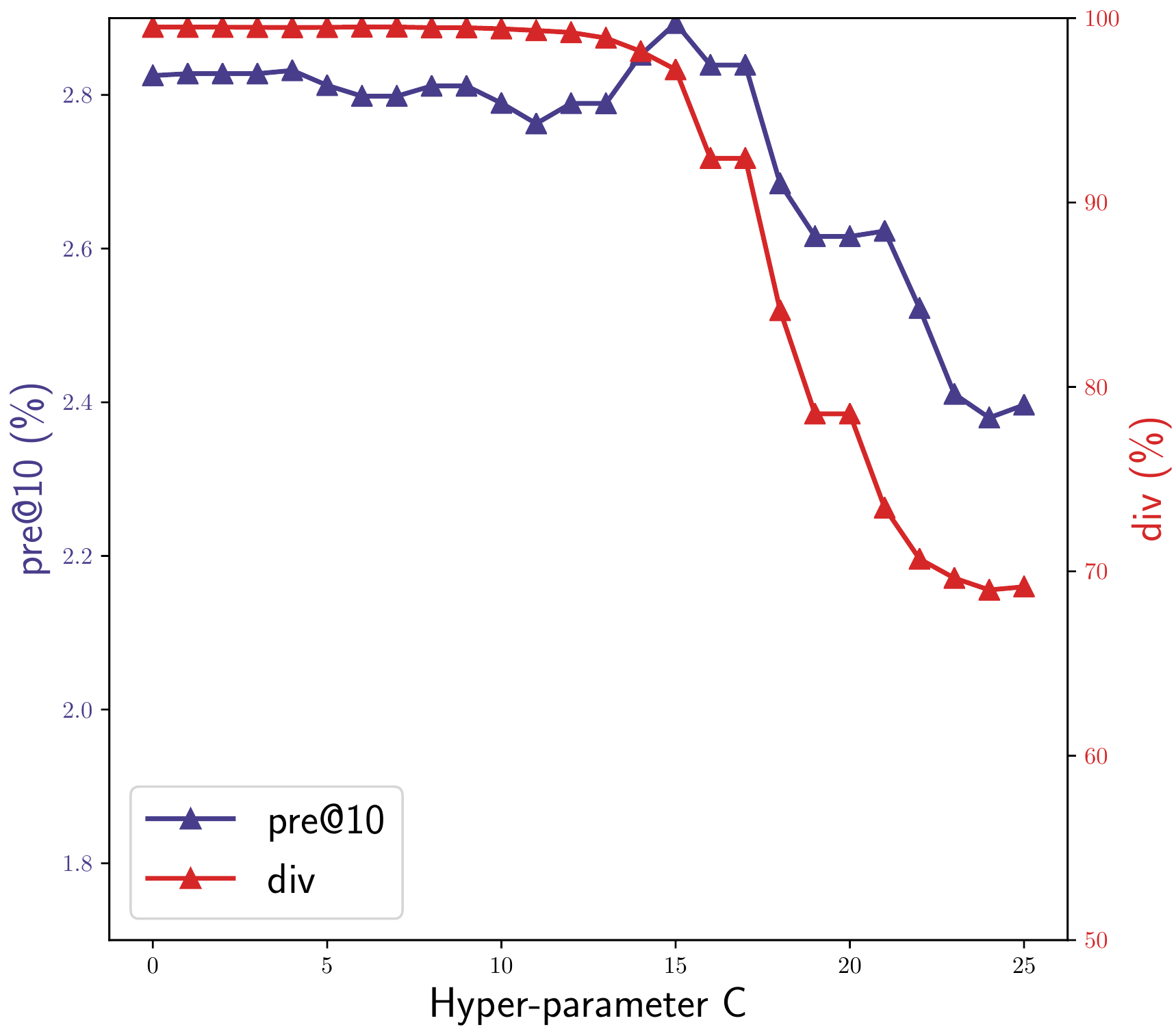}
			\label{fig:bsC}
		}
	\end{minipage}
	\begin{minipage}[t]{0.495\linewidth}
		\centering
		\subfigure[Bundle Size]{
			\includegraphics[width=1.65in]{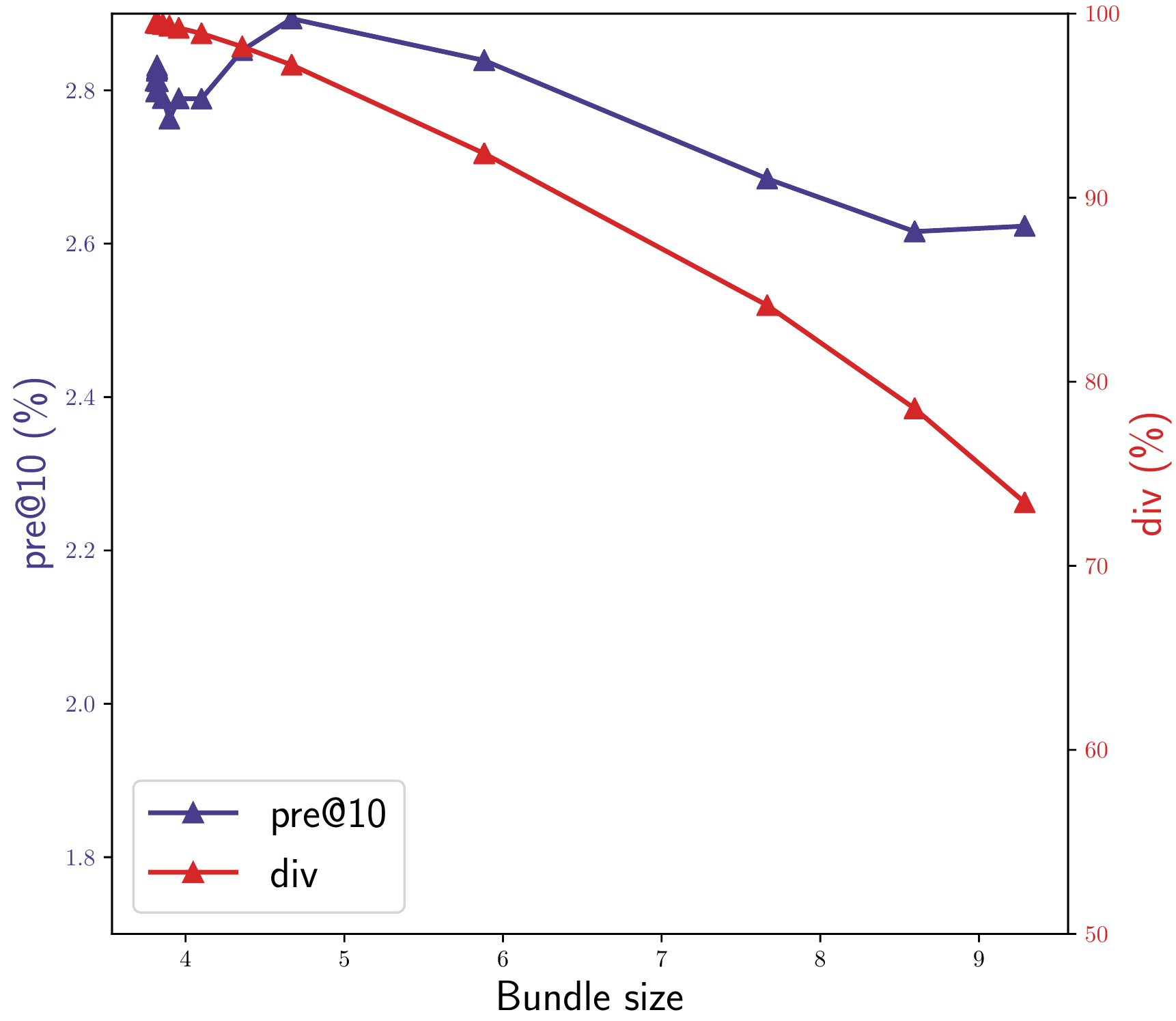}
			\label{fig:bsS}
		}
	\end{minipage}
	\vspace{-0.2cm}
	\caption{Performance Impact of the Masked Beam Search (Taobao)}
	\label{fig:masked_beam_search_p_div}
\end{figure}

\begin{figure*}[htbp]
	\centering
	
	\subfigure[Groundtruth]{
		\includegraphics[width=1.2in]{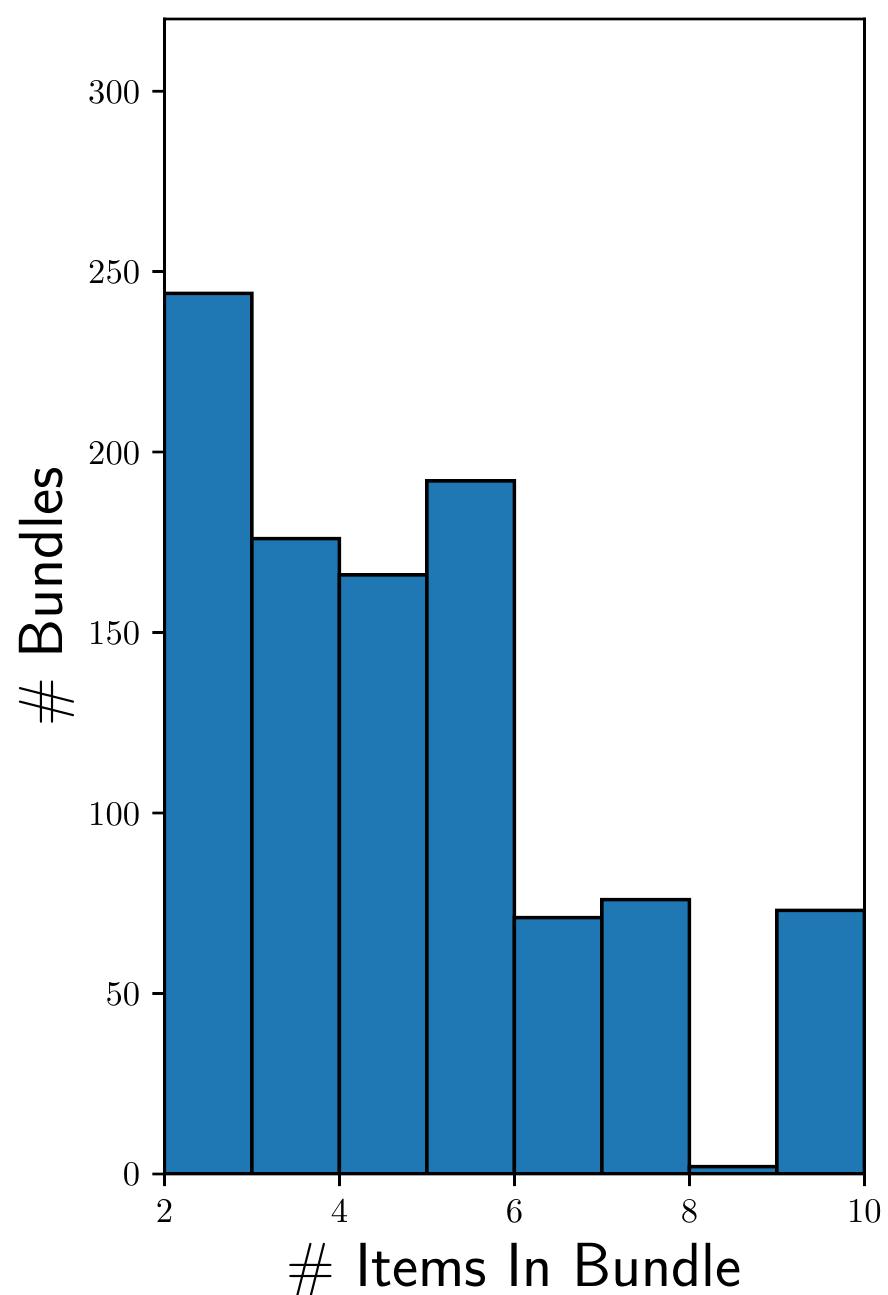}
		\label{fig:raw}
	}
	\subfigure[Original Beam Search] {
		\includegraphics[width=1.2in]{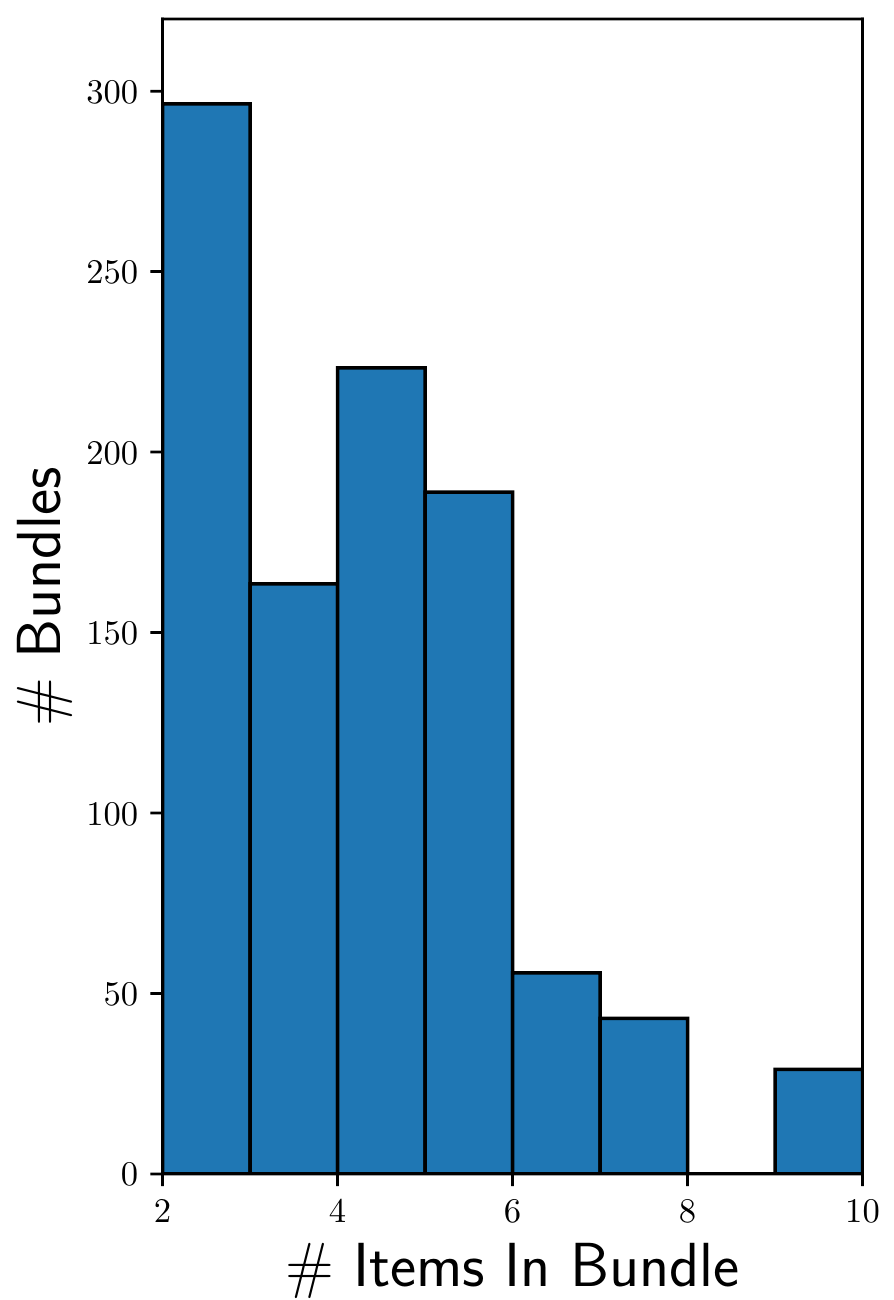}
		\label{fig:bs0}
	}
	\subfigure[C=14] {
		\includegraphics[width=1.2in]{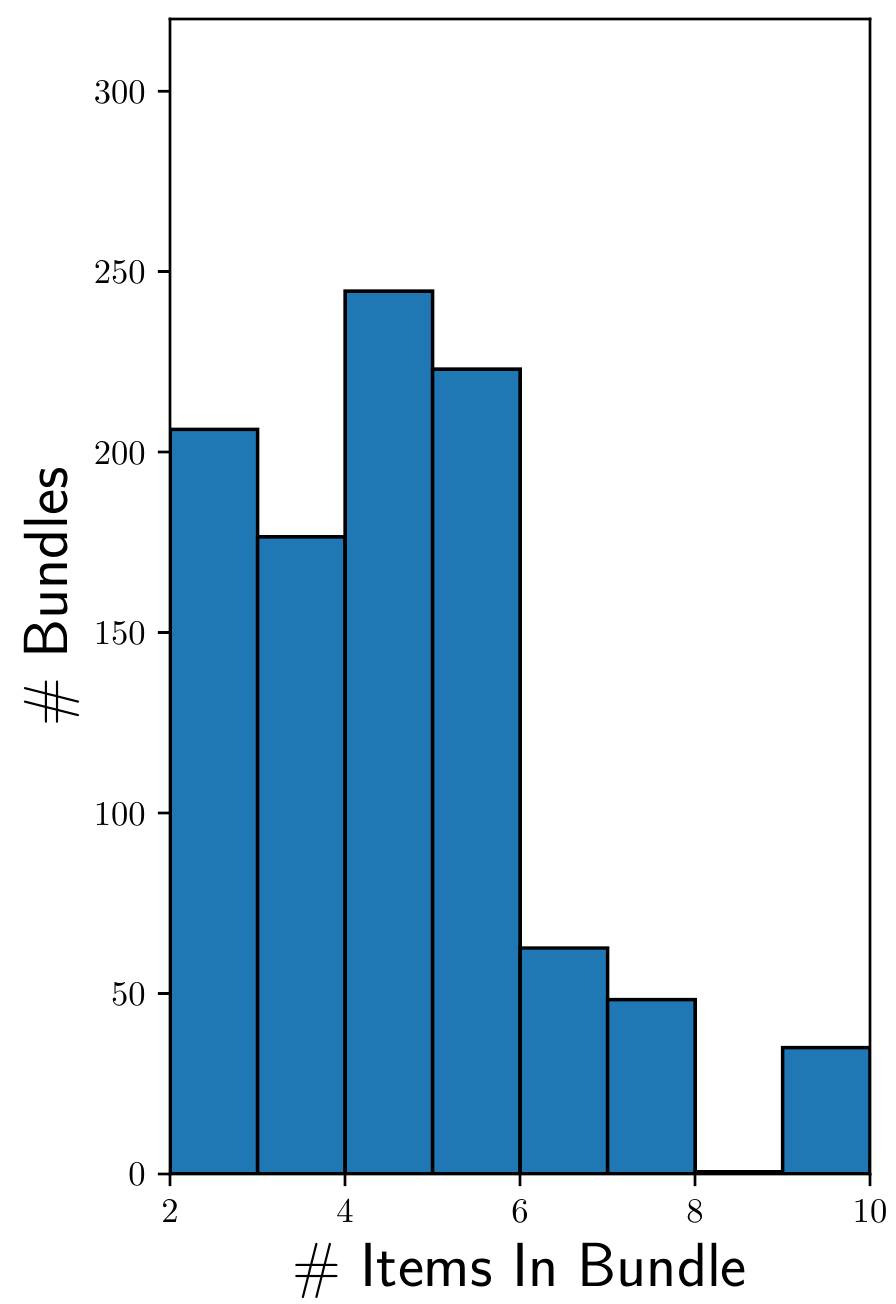}
		\label{fig:bs14}
	}
	\subfigure[C=17] {
		\includegraphics[width=1.2in]{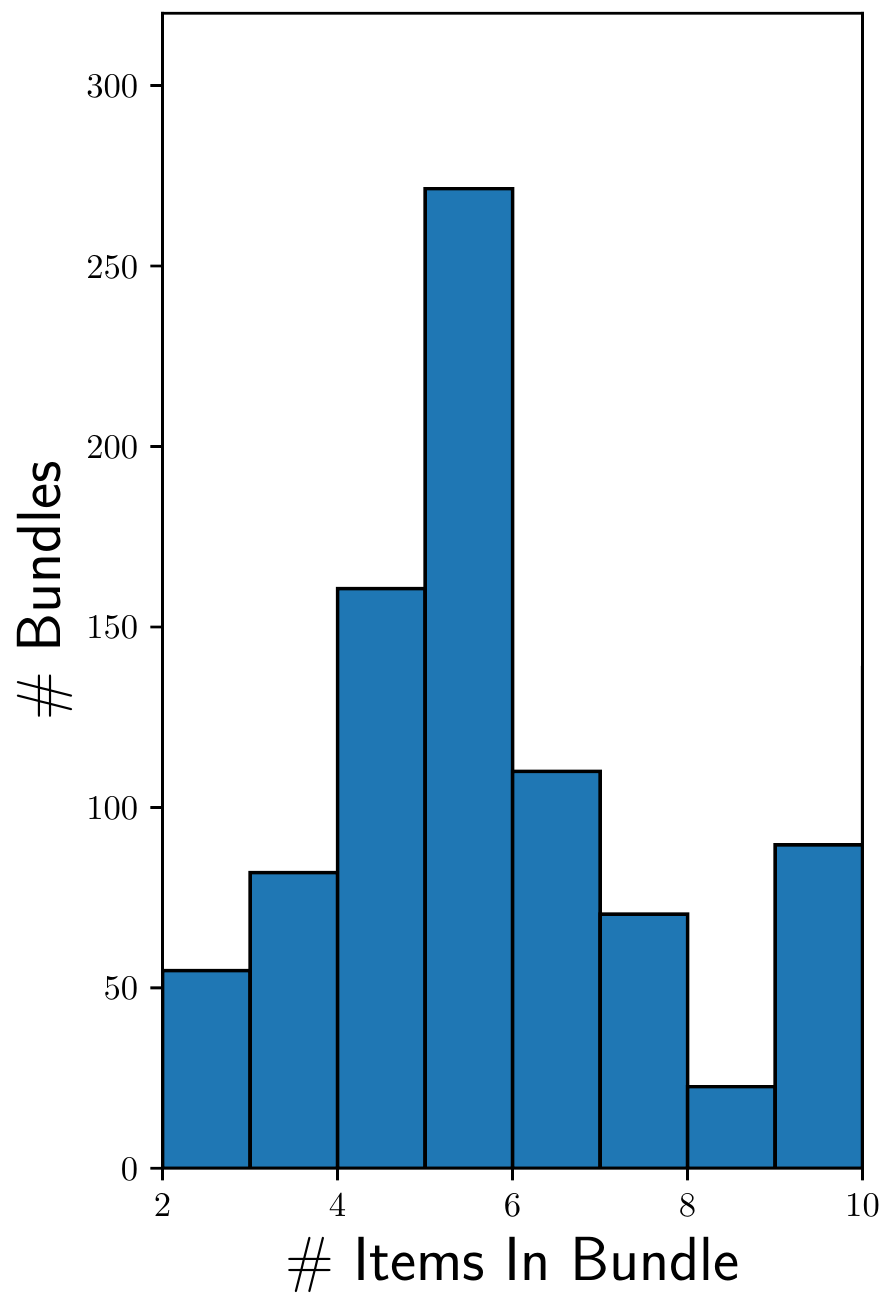}
		\label{fig:bs17}
	}
	\subfigure[C=20] {
		\includegraphics[width=1.2in]{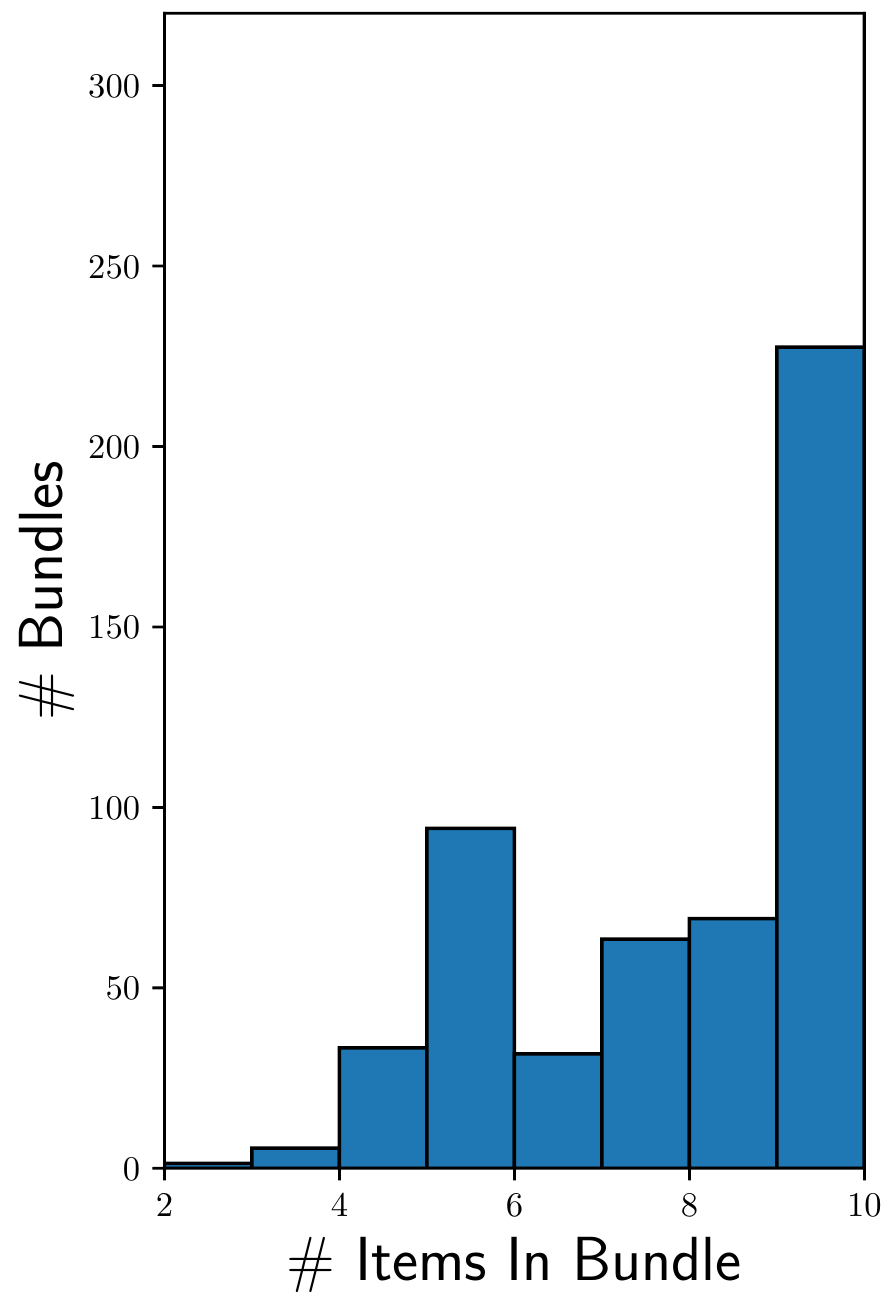}
		\label{fig:bs20}
	}
	\vspace{-0.3cm}
	\caption{Bundle Size Distribution Shifting by the Masked Beam Search (Taobao)}
	\label{fig:masked_beam_search_distribution}
	\vspace{-0.3cm}
\end{figure*}

The masked beam search controls the distribution of bundle size by changing a shifting typer-parameter $C$ according to Equation (\ref{eqa: shifting_parameter}), which reduces the unnormalized probabilities exponentially of generating the end token. We show in the Taobao dataset how the distribution of bundle size is altered in Figure \ref{fig:masked_beam_search_distribution} and the effect on the precision and diversity in Figure \ref{fig:masked_beam_search_p_div}.

In Figure \ref{fig:masked_beam_search_distribution}, we notice that the masked beam search does shift the distribution of bundle size. The distribution shifts significantly compared to the original beam search when setting C to be 14, and with the increase of C the average bundle size shifts as we expected. However, the shifting parameter $C$ enforces the network to generate the most likely items, although the generation should have ended. As shown in Figure \ref{fig:masked_beam_search_p_div}, $C$ is set from 0 to 20 with step 1.0, which leads to the bundle size varying from 3.8 to 9.2, then we show the performance impact about the precision and diversity respectively for the $C$ and bundle size. We notice that the precision still holds when we shift the bundle size from 3.8 to 6.0, but both the precision and diversity decrease if we continue shifting. Though our shifting strategy is simple, it is good enough to control the distribution in a small range in practice and it might be promoted in the future work through considering the long-term reward in advance by reinforcement learning.

\subsection{Balance Between Precision and Diversity}

\begin{figure*}[bp]
	\begin{minipage}[t]{0.245\linewidth}
		\centering
		\subfigure[Electro] {
			\includegraphics[scale=0.37]{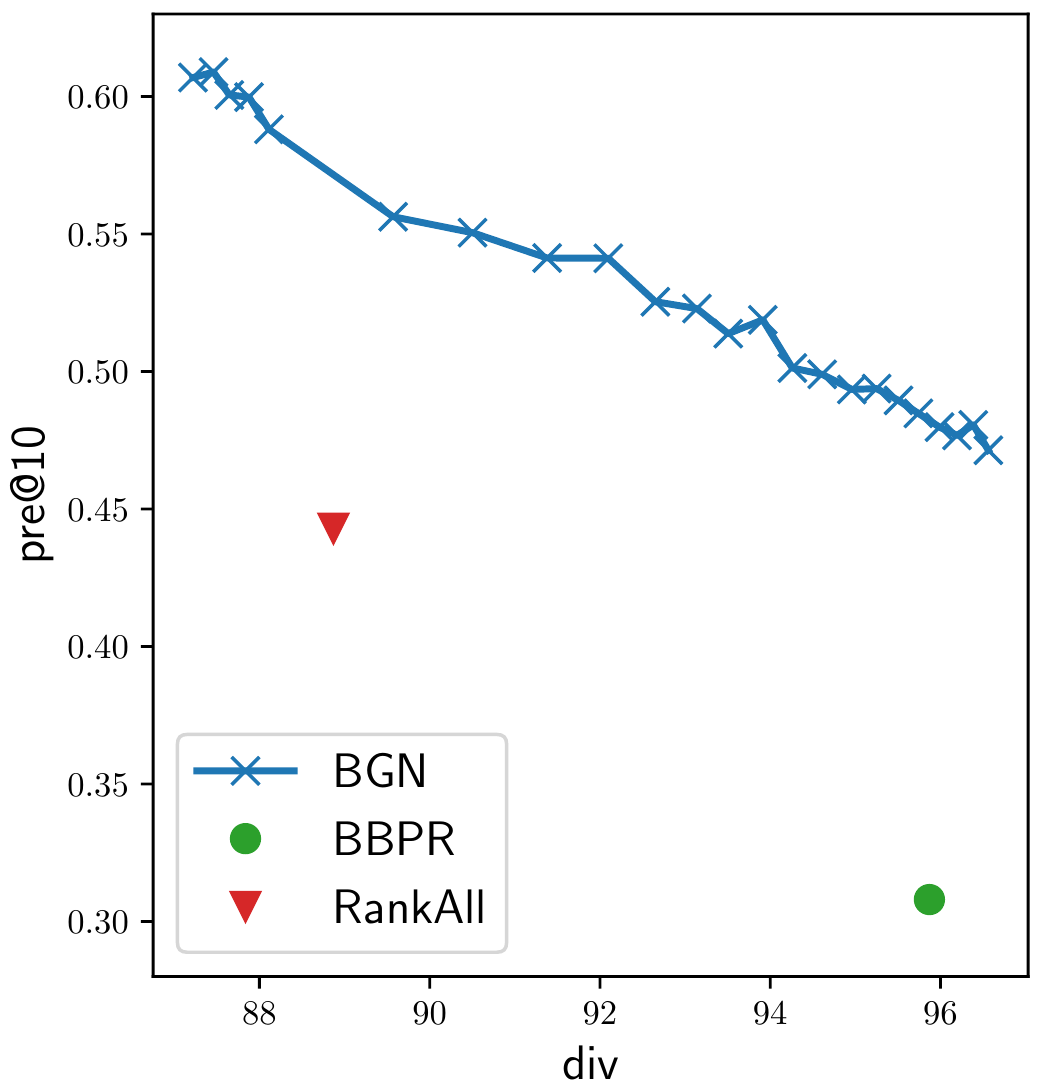}
			\label{fig:pdiv_clo}
		}
	\end{minipage}
	\begin{minipage}[t]{0.245\linewidth}
		\centering
		\subfigure[Clothe] {
			\includegraphics[scale=0.37]{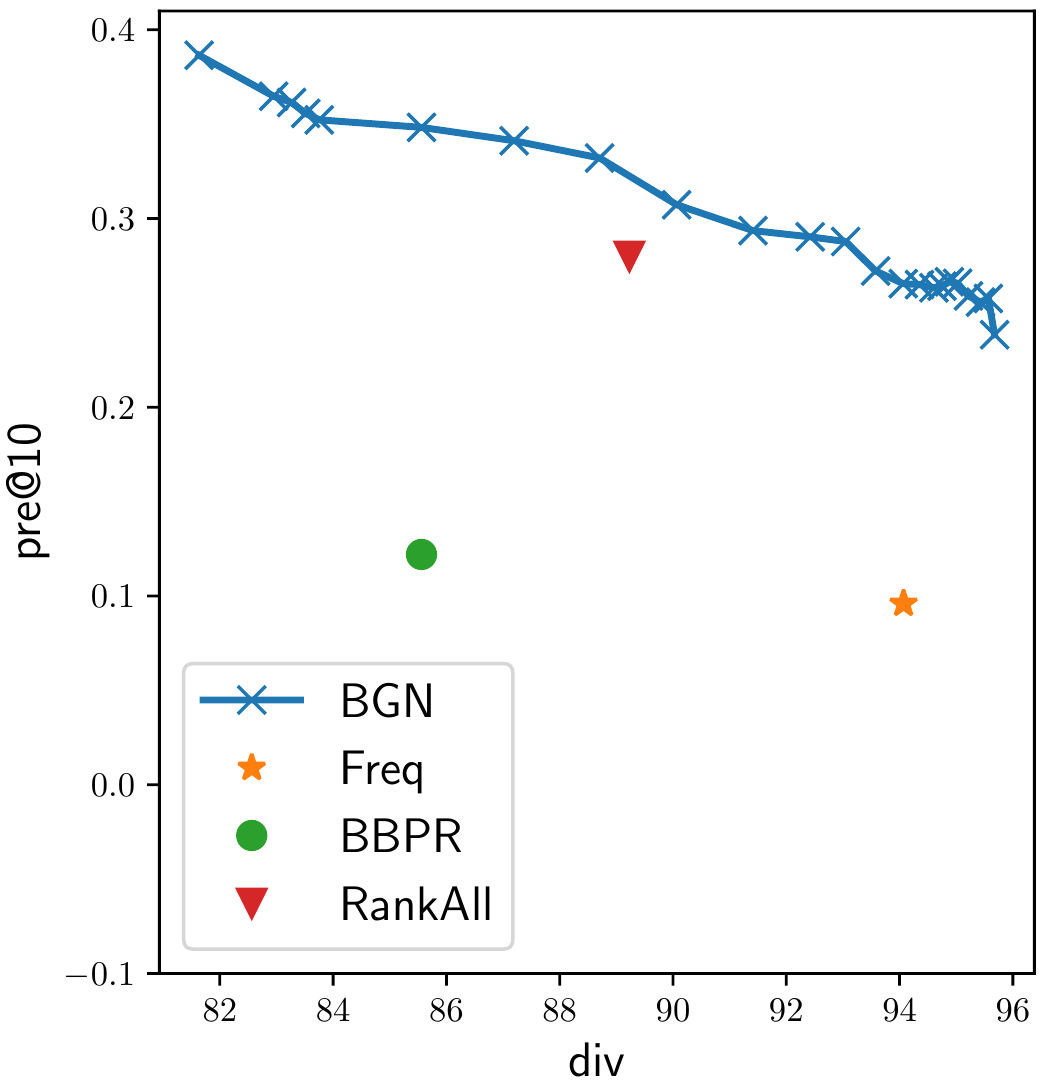}
			\label{fig:pdiv_ele}
		}
	\end{minipage}
	\begin{minipage}[t]{0.245\linewidth}
		\centering
		\subfigure[Steam] {
			\includegraphics[scale=0.37]{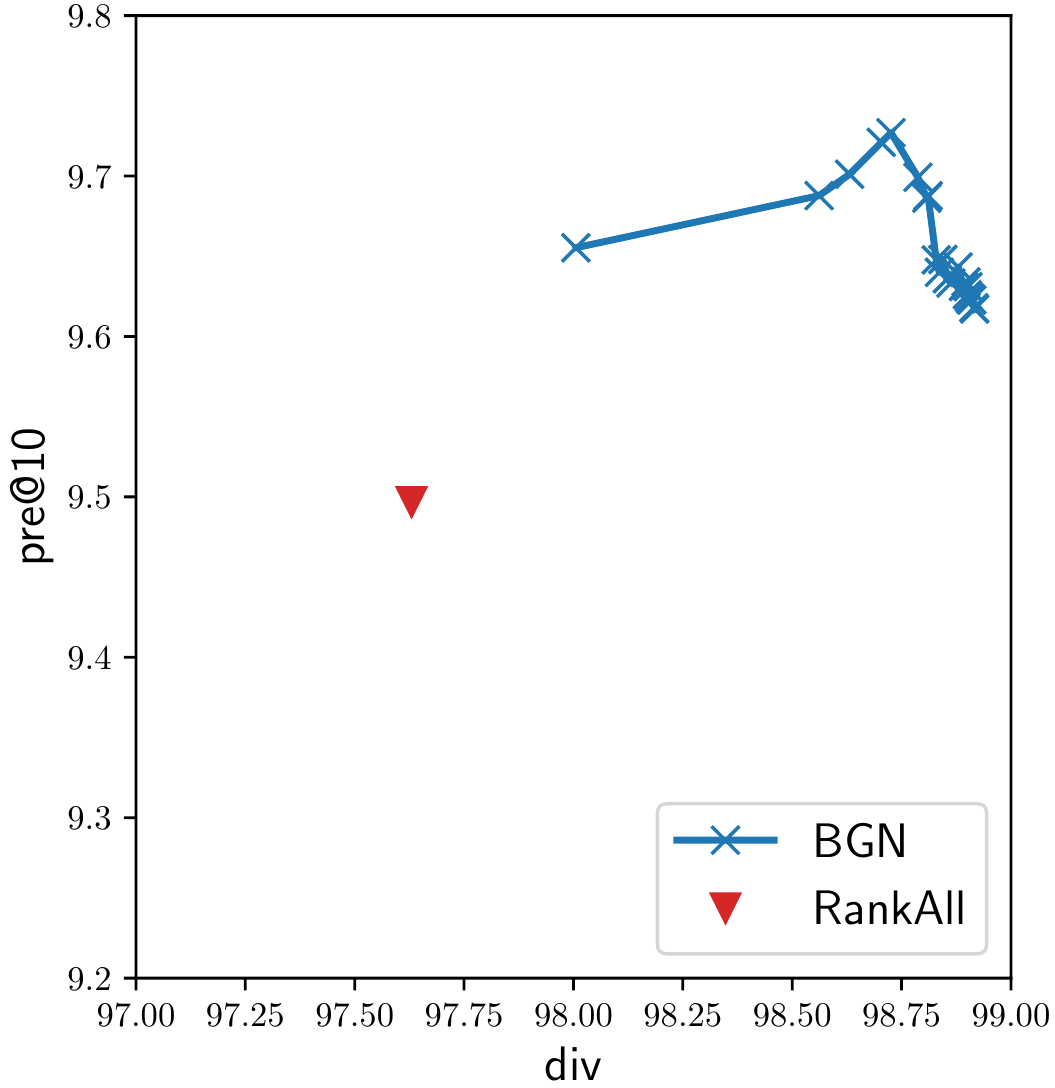}
			\label{fig:pdiv_ele}
		}
	\end{minipage}
	\begin{minipage}[t]{0.245\linewidth}
		\centering
		\subfigure[Taobao] {
			\includegraphics[scale=0.37]{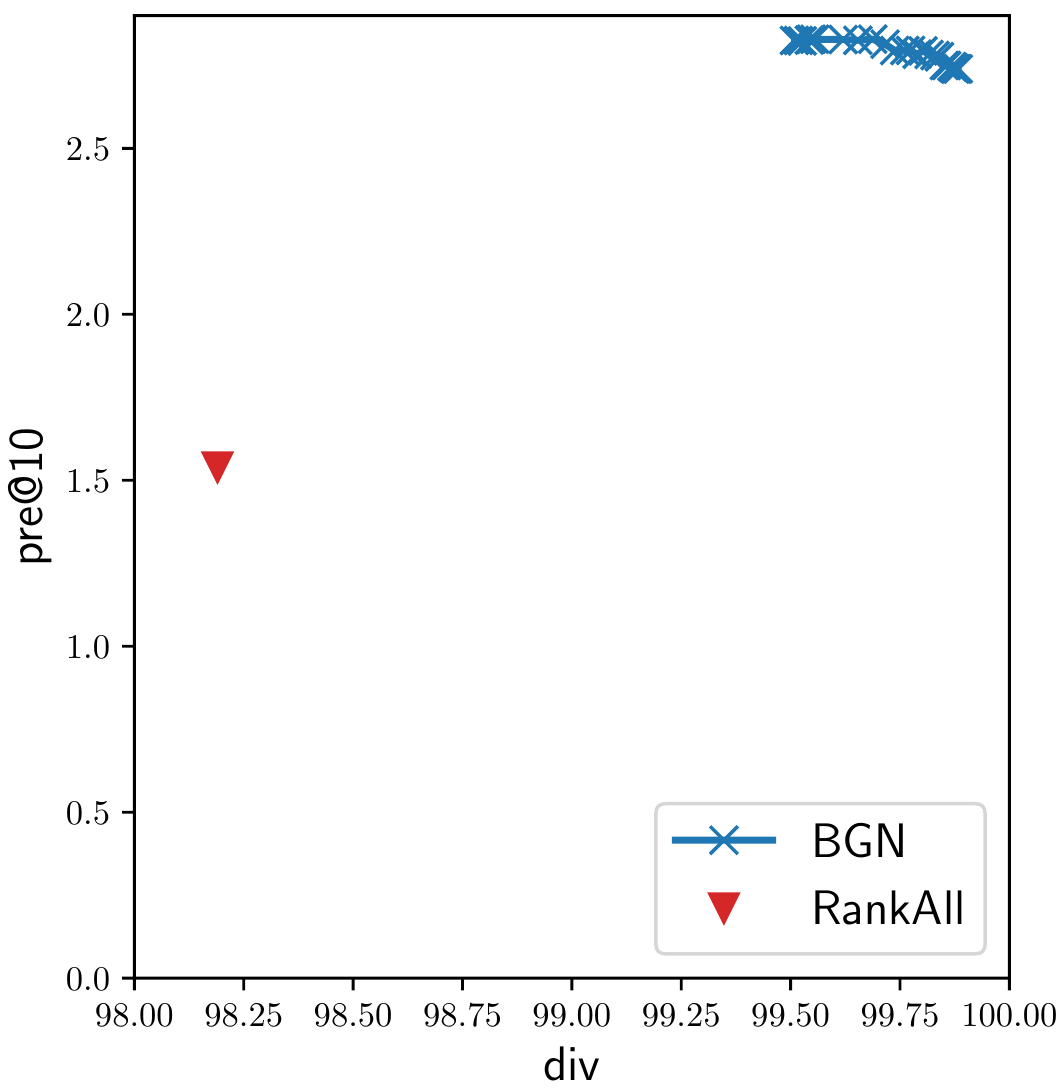}
			\label{fig:pdiv_ele}
		}
	\end{minipage}
	\vspace{-0.2cm}
	\caption{Balance between Precision and Diversity.}
	\label{fig:pdiv}
\end{figure*}

According to Algorithm \ref{algo: greedy}, we produce the diversified bundle list by combining the beam search and DPPs, which chooses the item maximizing the objective function of the quality and diversity in Equation (\ref{eqa: greedy_map}). The hyper-parameter $\lambda$ balances between the precision and the diversity. The metric of diversity is a separate metric from CTR and precision, trying to quantify the user-experience in a bundle recommendation system. When we increase $\lambda$, we pay more attention to the diversity among the generated bundles than the precision of the final list, and vice versa. We select $K$ bundles from $M$ candidates generated from the beam search with $M=50$, $K=10$ at each step, and $\lambda$ is set from 0 to 5 with step 0.25 and $C$ is set to be $0$.

We show in Figure \ref{fig:pdiv} that in all datasets, the precision decreases along with the growth of diversity as we expected. We also mark the other competitors' results as points in the figure, and we omit the competitors whose precision is too poor. Feature-aware softmax and beam search guarantee that BGN has a high precision base, meanwhile, DPP selection ensures to improve the diversity steadily with a tolerable reduction in precision. With adjustment of the $\lambda$, the curve formed by BGN is completely at the top right of all competitors. We achieve the highest precision and diversity in contrast with all competitors in all datasets.

\subsection{Comparison in the Bundle Generation Problem}

By comparing the proposed BGN against the other competitors including, \emph{Freq}, \emph{BRP}, \emph{BBPR}, \emph{RankAll}, \emph{Seq2seq}, we report and analyze their performance in terms of the precision metric $\textit{pre@k}$, the diversity metric $div$, and the average response time per user of these methods during inference, denoted by $avg\textrm{-}time$ in Table \ref{table:generation_pre_diversity}. Note that the shifting hyper-parameter $C$ of BGN is set to be $0$ in this part becasue we have shown the performance results of shifting the bundle size in Figure \ref{fig:masked_beam_search_p_div}.

\begin{table*}[htbp] 
	\centering
	\begin{tabular}{|c|c|c|c|c|c|c|c|c|c|c|c|c|}
		\hline
		& \multicolumn{6}{c|}{Synthetic Bundle} & \multicolumn{6}{c|}{Real bundle} \\ \cline{2-13} 
		Dataset & \multicolumn{3}{c|}{Electro} & \multicolumn{3}{c|}{Clothe} & \multicolumn{3}{c|}{Steam} & \multicolumn{3}{c|}{Taobao} \\ \cline{2-13} 
		& \textit{pre@10} & \textit{div} & \textit{avg-time} & \textit{pre@10} & \textit{div} & \textit{avg-time} & \textit{pre@10} & \textit{div} & \textit{avg-time} & \textit{pre@10} & \textit{div} & \textit{avg-time} \\ \hline
		Freq & 0.260\% & 95.56\% & - & 0.096\% & 94.07\% & - & 5.831\% & 69.44\% & - & 0.192\% & 60.00\% & - \\
		BRP & 0.324\% & - & 4.85s & 0.149\% & - & 3.96s & 6.750\% & - & 0.44s & 0.332\% & - & 5.47s \\
		BBPR & 0.308\% & 95.87\% & 0.68s & 0.122\% & 85.56\% & 0.34s & 7.168\% & 87.32\% & 0.29s & 0.357\% & 95.49\% & 1.27s \\
		RankAll & 0.443\% & 88.87\% & 55.97s & 0.280\% & 89.23\% & 7.94s & 9.497\% & 97.63\% & 0.11s & 1.540\% & 98.19\% & 1.38s \\
		\eat{Seq2seq & 0.29\% & 85.94\% & \textbf{0.04s} & 0.10\% & 75.93\% & \textbf{0.03s} & 9.32\% & 98.25\% & \textbf{0.02s} & 1.726\% & 99.77\% & \textbf{0.13s} \\} \hline
		BGN ($\lambda=0$) & \textbf{0.607\%} & 87.22\% & \textbf{0.05s} & \textbf{0.387\%} & 81.63\% & \textbf{0.03s} & \textbf{9.655\%} & 98.01\% & \textbf{0.04s} & \textbf{2.825\%} & 99.50\% & \textbf{0.24s} \\
		BGN ($\lambda=5$) & 0.469\% & \textbf{96.68\%} & 0.09s & 0.235\% & \textbf{95.75\%} & 0.06s & 9.617\% & \textbf{99.08\%} & 0.06s & 2.739\% & \textbf{99.88\%} & 0.28s \\ \hline
	\end{tabular}
	\vspace{0.1cm}
	\caption {$pre@10$, $div$, $avg\textrm{-}time$ of Different Methods in the Bundle List Recommendation Problem.}
	\label{table:generation_pre_diversity}
	\vspace{-0.5cm}
\end{table*}

\emph{Freq} performs poorly in term of precision on all datasets, because it is a non-personalized method, but \emph{Freq} performs quite diverse in Amazon dataset. We think that it is because the bundles of two Amazon datasets are generated synthetically by co-purchase records, which contain more diversified bundles naturally, whereas, the bundles of Steam and Taobao dataset is pre-defined by sellers.

\emph{BRP} produces only one bundle, so there is no diversity reported. We repeat the results K times to produce the list for achieving the higher precision, which is well-defined because our metric of precision averages on each position. With regard to precision, \emph{BRP} performs better than \emph{Freq} owing to the personalization and consideration of the the first and second cross-item dependencies, but it ignores the higher order terms and gives a sub-optimal result. Besides, compared with other methods, solving the mixed-integer quadratic programming (MIQP) problem is time-consuming.

\emph{BBPR} produces a personalized list of bundles instead of one. This method utilizes both item and bundle information including bundle size and the pair-wise Pearson correlation of the items, which performs better than \emph{Freq} and \emph{BRP} on most of the dataset. However, the annealing schedule leads to randomness in generating bundles and is easily caught in the local optimum due to instability of annealing.

\emph{RankAll} performs better than \emph{BBPR} and \emph{BRP} on the metrics of precision and diversity on all datasets, because it uses CNN as a context extractor capturing more combinatorial generalization of features. But for the response time, \emph{RankAll} is the most time-consuming approach on almost all datasets. Because the inference time of \emph{RankAll} is approximately proportional to the number of bundles, whereas, other methods are approximately proportional to the number of items. Essentially, \emph{RankAll} is not a generative model, but a ranking model which ranks the existing bundles to give the final recommendation list. It wastes most of the time scoring the low-quality bundles and it is not scalable in practice. Furthermore, it could not produce new bundles beyond the existing ones in training data, which limits the precision.

\eat{\emph{Seq2seq} uses the encoder-decoder framework and applies two layers' stacked LSTM cell. Besides the encoder where BGN uses the text-CNN, the major difference of \emph{Seq2seq} from BGN is that there is no feature-aware softmax and diversified process for the regular \emph{Seqseq}. We notice that \emph{Seq2seq} performs fairly in the real bundle dataset, but it can not handle the synthetic bundles well in the Amazon dataset because the co-purchase items are more randomized. \emph{Seq2seq} has the fastest inference time, whereas BGN costs extra time on the feature-aware softmax and the DPP selection.}

\emph{BGN} achieves the best precision, diversity and fastest response time at the same time on all datasets. The high precision benefits from the the feature-aware softmax. However, when setting $\lambda$ to be $0$, BGN has low diversity especially in the synthetic bundle because traditional seq2seq is easy to generate similar sentences. When we simply set the $\lambda$ to be $5$, BGN balances the quality and diversity, which improves the precision of the best baselines by 16\% on average while maintaining the highest diversity on four datasets. Besides, BGN is a sequnce generation approach without ranking or recursive searching. The time complexity of BGN during inference is proportional to the number of items. So our model achieves fastest response time in bundle list recommendation problem. Specifically, BGN improves the response time by 6.5x, 4.6x, 0.8x, 3.5x on four datasets respectively over the best competitors when $\lambda$ equals $5$.

\subsection{Comparison in the Bundle Ranking Problem}

BGN is designed for the bundle list generation problem. However, we can adopt it in the bundle ranking problem, by feeding the candidate bundles to the inputs of the decoder and taking the negative loss as the ranking score, so that we compare with other competitors in the bundle ranking problem. Note that, the absolute values of the AUC in the bundle ranking problem are usually lower than the one in the item ranking problem because the sample space for bundles is larger and more sparse.

\begin{table}[htp] 
	\centering
	\begin{tabular}{|c|cccc|}
		\hline
		AUC & Electro & Clothe & Steam & Taobao \\ \hline
		BBPR & 74.85\% & 66.20\% & 90.27\% & 60.24\% \\
		\eat{\multicolumn{1}{|l|}{Seq2seq} & 74.63\% & 66.82\% & 98.78\% & 60.74\% \\}
		RankAll & 75.51\% & 66.97\% & \textbf{99.66\%} & 60.93\% \\
		BGN & \textbf{78.11\%} & \textbf{69.99\%} & 99.36\% & \textbf{61.27\%} \\ \hline
	\end{tabular}
	\vspace{0.1cm}
	\caption {AUC Tests in the Bundle Ranking Problem.}
	\label{table:AUC}
	\vspace{-0.5cm}
\end{table}

In Table \ref{table:AUC}, we show the measurement of AUC for different methods, and BGN improves the best baseline by 2.04\% on average on four datasets. Actually, in the bundle ranking setup, all competitors in Table \ref{table:AUC} can be regarded as the encoder-decoder framework, where the encoder extracts the user context information and the decoder extracts the bundle information. The major differences are the architectures of specific neural network and the loss function. With the help of the feature-aware softmax, BGN is good at utilizing the feature information, because the weight matrix of softmax is built dynamically and its loss is a generalization of optimizing the average BPR at each step. The Steam dataset has no category feature, so BGN is worse than \emph{RankAll} slightly on this dataset.

\section{Conclusions}
This paper studies a problem of personalized bundle list recommendation and proposes a bundle generation network (BGN). BGN decomposes the problem into the quality/diversity part to produce the high-quality and diversified bundle list. The proposed feature-aware softmax makes up the inadequate representation of traditional sequence generation for rich features in bundle recommendation scenario, which improves the modeling of quality significantly. We conduct extensive experiments on three public datasets and one industrial dataset. BGN achieves the best state-of-the-art results on the metrics of precision, diversity and response time on all datasets. Besides, BGN can control the bundle size with a tolerable reduction in precision.

\begin{acks}
	This work was partially supported by \grantsponsor{}{National Key Research and Development Program}{} No. \grantnum{}{2016YFB1000700},  \grantsponsor{}{NSFC}{} under Grant No. \grantnum{}{61572040} and \grantnum{}{61832001}, and \grantsponsor{}{Alibaba-PKU joint Program}{}.
\end{acks}

\balance
\bibliographystyle{ACM-Reference-Format}
\bibliography{main}

\end{document}